\documentclass[10pt]{article}

\usepackage[english]{babel}
\usepackage[utf8]{inputenc}
\usepackage[T1]{fontenc}
\usepackage[]{todonotes}

\usepackage{url}
\usepackage{booktabs}
\usepackage{lscape}
\usepackage{longtable}
\usepackage{tabularx}
\usepackage{xspace}
\usepackage{multirow}
\usepackage{float}
\usepackage{array}
\usepackage{graphicx}
\usepackage{pbox}
\usepackage{makecell}
\usepackage{color}
\usepackage{colortbl}
\usepackage{csvsimple}

\usepackage{mdframed}
\usepackage{xcolor}
\usepackage{eurosym}

\usepackage{flushend}

\usepackage{enumitem}

\usepackage[multiple]{footmisc}

\usepackage{lineno}

\usepackage{footmisc}

\usepackage{fancyhdr}
\usepackage{hyperref}
\usepackage[nottoc]{tocbibind}

\setitemize{noitemsep,topsep=0pt,parsep=0pt,partopsep=0pt}
\setlist[itemize]{leftmargin=*}

\newcounter{seagileinsight}

\newcommand{\problemref}[1]{\color{magenta}\textit{(#1)}\color{black}\xspace}
\newcommand{\practiceref}[1]{\color{teal}\textit{(#1)}\color{black}\xspace}

\newcommand{\coderef}[2]{\textsc{\small{#1 (\textsl{g}=#2)}}}

\newcommand{\groundedsym}{\textsl{g}}

\begin{document}
	
	\pagestyle{empty}
	
	\title{System and Software architecting harmonization practices in ultra-large-scale Systems of Systems}

	\author{
		H\'ector Cadavid\\
		\texttt{h.f.cadavid.rengifo@rug.nl}
		\and
		Vasilios Andrikopoulos\\
		\texttt{v.andrikopoulos@rug.nl}
		\and
		Paris Avgeriou\\
		\texttt{p.avgeriou@rug.nl}
		\and
		P. Chris Broekema\\
		\texttt{broekema@astron.nl}
	}
	
	\date{}
	
	\maketitle

\begin{abstract}

\textbf{Context:} The challenges posed by the architecting of System of Systems (SoS) has motivated a significant number of research efforts in the area. However, literature is lacking when it comes to the interplay between the disciplines involved in the architecting process, a key factor in addressing these challenges. \textbf{Objective:} This paper aims to contribute to this line of research by confirming and extending previously characterized architecting harmonization practices from Systems and Software Engineering, adopted in an ultra-large-scale SoS. \textbf{Method:} We conducted a confirmatory case study on the Square-Kilometre Array (SKA) project to evaluate and extend the findings of our exploratory case on the LOFAR/LOFAR2.0 radio-telescope projects. In doing so, a pre-study was conducted to map the findings of the previous study with respect to the SKA context. A survey was then designed, through which the views of 46 SKA engineers were collected and analyzed. \textbf{Results:} The study confirmed in various degrees the four practices identified in the exploratory case, and provided further insights about them, namely: (1) the friction between disciplines caused by long-term system requirements, and how they can be ameliorated through intermediate, short-term requirements; (2) the way design choices with a cross-cutting impact on multiple agile teams have an indirect impact on the system architecture; (3) how these design choices are often caused by the criteria that guided early system decomposition; (4) the seemingly recurrent issue with the lack of details about the dynamic elements of the interfaces; and (5) the use of machine-readable interface specifications for aligning hardware/software development processes.  \textbf{Conclusions:} The findings of this study and its predecessor support the importance of a cross-disciplinary view in the Software Engineering research agenda in SoS as a whole, not to mention their value as a convergence point for research on SoS architecting from the Systems and Software Engineering standpoints.

\end{abstract}

\pagestyle{fancy}

\thispagestyle{empty}

\section{Introduction}
The concept of System of Systems (SoS) is extensively used in application domains like defense~\cite{dod2003dod}, manufacturing~\cite{alfieri2012sos,rauschecker2014developing}, emergency systems~\cite{ki2018system,zachry2018establishing}, energy \cite{steghofer2013system,lopes2011model}, and  health care~\cite{gorod2018toward,okami2017transitional} to describe \textit{a collection of systems that cooperate to fulfill a goal or to  provide new capabilities}~\cite{ISO21839}. By definition, most SoS, like those in the aforementioned domains, are characterized by different degrees of \textit{Managerial and Operational Independence}, \textit{Diversity}, \textit{Heterogeneity}, \textit{Emergence} and \textit{Belonging} as proposed by Maier~\cite{maier1998architecting} and Boardman et al.~\cite{boardman2006report}. Such characteristics make SoS a powerful concept, but at the same time pose significant challenges to their architecting process. For example, the dynamic nature of an SoS  makes it difficult to anticipate its behavior at design time, and therefore quality requirements are difficult to address~\cite{ceccarelliBasicConceptsSystems2016,gagliardi2009uniform}. 

Although there have been significant research efforts to address these challenges~\cite{cadavid2019architecting}, there is still little research on a key element of the SoS architecting process: the interplay between the disciplines involved in it, namely Systems Engineering and Software Engineering. To contribute towards addressing this research gap, we have conducted a series of studies which, together, aimed at: (1) identifying and characterizing the pain points of the software architecting processes in large-scale SoS where development is usually driven by Systems Engineering (SE)~\cite{muller2012validation}; and (2) identifying and characterizing the best practices to harmonize said architecting processes. The first study of the series~\cite{cadavid2020survey} confirmed that major integration and operational problems in SoS are indeed linked to the way different disciplines work together on their architecting process. Particular instances of such problems ---related to requirements and interfaces--- were identified. In the second study~\cite{cadavid2021casestudy}, we identified a number of best practices that engineers and architects have adopted exactly to address the aforementioned interdisciplinary issues through an exploratory case study on a long-running, large-scale SoS:
the LOFAR radio telescope and its follow-up project, LOFAR2.0\footnote{\url{https://www.astron.nl/telescopes/lofar/}}. These practices concern (1) how to define high-level requirements and make them properly propagate to the lower levels (e.g., as software requirements), (2) how to properly demarcate the boundaries of responsibilities at the lower levels of the system through system decomposition and high-level interfaces (3) how to organize the architecting roles of the disciplines involved, and (4) how to achieve early integration in the process. 

In the current paper, we present a follow-up study that aims \emph{to confirm and extend our previous findings} with respect to architecting practices and the cross-disciplinary issues they address. For this purpose we conduct a confirmatory case study on a ultra-large-scale SoS: the Square Kilometre Array\footnote{\url{https://www.skatelescope.org/the-ska-project/} The reader is advised that information about the SKA presented in this study is valid at the time of writing (November 2021). As with any ongoing project, this is potentially subject to change in the future.} (SKA). The SKA is a global project, with eleven international consortia involved in its design, that aims to build multi-purpose low-band/mid-band radio telescopes in South Africa and Western Australia which, when working together, will collect data from an area equivalent of (at least) one square kilometre. 

The overall contribution of this work can be summarized as follows. First, it confirmed in various degrees the four practices identified in the exploratory case study~\cite{cadavid2021casestudy}, namely a) \textit{Adopting a rigorous systems-engineering process where requirements are defined upfront}; b) \textit{Hierarchical co-architecting roles}; c) \textit{Splitting up the system on a subsystem level, with boundaries demarcated with \textit{Interface Control Documents (ICDs)}}; and d) \textit{Early Integration activities/techniques}. Second, it provided further insights on these practices, in particular: (a) the friction between disciplines caused by the long-term system requirements ---and hence the call for intermediate, properly traceable short-term requirements derived from them; and (b) how some prescriptive design decisions (with implications across multiple teams), and the criteria used to guide the early system decomposition have an indirect impact on the system architecture and qualities. Third, it unveiled additional and seemingly successful practices to bridge the gaps between disciplines that range from cross-level communication practices, to more technical ones oriented towards the alignment of hardware and software development cycles.

The rest of this paper is structured as follows: Section~\ref{sec:background} summarizes the LOFAR/LOFAR2.0 study and its findings, and the particular setting of the SKA. Section~\ref{sec:study-design} presents the study design. Section~\ref{sec:results} describes the study results, and Section~\ref{sec:discussion} discusses the answers to the proposed research questions based on synthesizing these results. Section~\ref{sec:implications} discusses the implications of the findings for practitioners and researchers. Finally, Sections~\ref{sec:ttv} and~\ref{sec:conclusions} present the \textit{Threats to Validity} and the general conclusions of the study, respectively.

\section{Background}\label{sec:background}
This section briefly describes the main findings of the previous case study and the characteristics of the case subject, the SKA system, from which said findings are expected to be confirmed and expanded. A full description of the former is available at~\cite{cadavid2021casestudy}.

\subsection{LOFAR+ exploratory case study \& its identified practices}

LOFAR (Low Frequency Array) is one of the largest radio telescopes on Earth, and falls into the category of \textit{directed SoS}, i.e., an SoS built and centrally managed to fulfill specific purposes~\cite{maier1998architecting,Dahmann2008}. It consists of multiple, geographically distributed stations, which in turn, operate massive antenna arrays whose readings are digitized, filtered, and transported at a central location to be processed according to the requirements of a number of scientific data products~\cite{van2013lofar}. As many SoS, its development occurred in a particularly long time frame (over two decades), and involved numerous scientific and engineering challenges that led to many lessons learned. Consequently, LOFAR2.0, the ongoing expansion of the scientific and technical capabilities of LOFAR (expected to be ready in 2025), relies on the experience gathered from its predecessor~\cite{Hessels:2016}. Our previous exploratory case study on both projects~\cite{cadavid2021casestudy} (from now on referred to as the \textit{LOFAR+ case}) focused on characterizing the practices adopted as a consequence of this experience towards harmonizing the architecting practices of the disciplines involved in the project, and in particular between SE and SWE.

Four high-level practices were identified, namely: \textit{Adopting a rigorous systems-engineering process where requirements are defined upfront}~\practiceref{B1}, \textit{Hierarchical co-architecting roles}~\practiceref{B2}, \textit{Splitting up the system on a subsystem level, with boundaries demarcated with \textit{Interface Control Documents (ICDs)}}~\practiceref{B3}, and \textit{Early Integration activities/techniques}~\practiceref{B4}. The cross-disciplinary issues identified in LOFAR and addressed in LOFAR2.0 through the aforementioned practices are related to: a) \textit{system requirements and how they flow down to the lower levels of the system}; b) \textit{local design decisions and how they may impact the overall system}; and c) \textit{subsystem interfaces, and how these capture the mutual expectations of the involved parties}. 
In the following, practices \practiceref{B1}-\practiceref{B4} are described in more detail, while their related issues, \problemref{P1}-\problemref{P9}, will be discussed further when discussing the study results in the following section:

\begin{description}[leftmargin=1em]
	\item[\practiceref{B1}] In LOFAR there were issues related to the system-level requirements (also known as L1-requirements), and they way they flowed down as subsystem (e.g., software) requirements. In particular, it was perceived  that in many cases the requirements were unclear or missing \problemref{P1}, arguably due to the fact that LOFAR, was originally proposed as an instrument concept that would take advantage of a technological opportunity, where its possibilities were yet to be discovered. It was perceived that this, and other problems such as not having properly defined interfaces \problemref{P6}, \problemref{P7}, was caused by the lack of a proper systems engineering process, where requirements are defined up-front (in other words, requirements front-loading). Consequently, on LOFAR2.0, this was seen as a key practice.

	\item[\practiceref{B2}] In LOFAR, the space for making design decisions on the subsystems (particularly the software-intensive ones) was not clearly demarcated by the system requirements~\problemref{P3}. 
	Given the lack of a project role that guards/drives the overall system qualities \problemref{P5}, this contributed, in turn, to the emergence of architectural smells from local design decisions~\problemref{P4}. Consequently, adopting hierarchical co-architecting and co-design approaches \practiceref{B2}, where system architects oversee both system and software subsystems/components, was seen as a key practice in LOFAR2.0.
	
	\item[\practiceref{B3}] In LOFAR, the requirements at software level were too constrained by the design decisions at system level \problemref{P2} as the software side was involved late in the process. Furthermore, the local design decisions under this particular setting lead to the emergence of architecture smells~\problemref{P4}. In LOFAR2.0, it was perceived that having the system split at the subsystem level early on (instead of a hardware-, software-, and firmware- based partition), with properly defined ICDs as boundaries, was key to address these issues.

	\item[\practiceref{B4}] In LOFAR, the integration process was considered challenging due to the substantial tweaking and trouble-shooting required. To a large extent, this is attributed to issues with the subsystem interfaces. In particular, it was perceived that such interfaces were not thoroughly thought-out \problemref{P6} and (arguably as a consequence of this), they did not sufficiently capture the mutual expectations between the two parties involved in them \problemref{P7}. For this reason, in LOFAR2.0 early integration activities/techniques \practiceref{B4} were adopted as a means to anticipate misunderstandings, and avoid wrong assumptions from both parties of an interface (e.g. hardware and software). 

\end{description}

It is worth noting that two more issues were identified, that were not solved when transitioning from LOFAR to LOFAR2.0. First, the integration of hardware, firmware and software was challenging given that the traditionally long development cycles of the hardware, and the short (agile) cycles of software were difficult to align \problemref{P8}. Second, the study revealed that the requirements were focused mostly on the steady-states of the system. That is to say, the dynamic behavior of the system and its constituents seemed to be overlooked \problemref{P9}. 

\subsection{SKA overview}\label{sec:SKA-overview}
The Square Kilometre Array is a hierarchical, distributed, ultra-large-scale next-generation radio telescope, which entered its construction phase in June of 2021. The construction budget for this phase of the telescope is around 700M\euro. Scientific operations of the first phase of this instrument, consisting of a 197 dish mid-frequency array (including 64 existing MeerKAT dishes) in the South African Karoo desert and a 130.000 element low-frequency phased array in Western Australia, is expected to start sometime around 2027. Total construction of the first phase of SKA, including contingency and labour, will cost around 1.28 billion \euro\footnote{SKA Phase 1 construction proposal (\url{https://www.skatelescope.org/key-documents/})}.

The SKA construction, commissioning and exploitation is governed by the SKAO, the SKA Observatory, a non-governmental organisation (NGO) that was established in March 2019 and came into effect in January 2021. The SKA design was conceived during a protracted  design phase, which saw contributions from sixteen countries and around a hundred different organisations. While no public estimate exists on the number of engineers involved in this process, project-wide engineering meetings consistently drew between two and three hundred engineers.

\subsubsection{SKA high-level architecture}\label{sec:SKA-architecture}

\begin{figure}[t]
	\centering
	\includegraphics[width=0.7\linewidth]{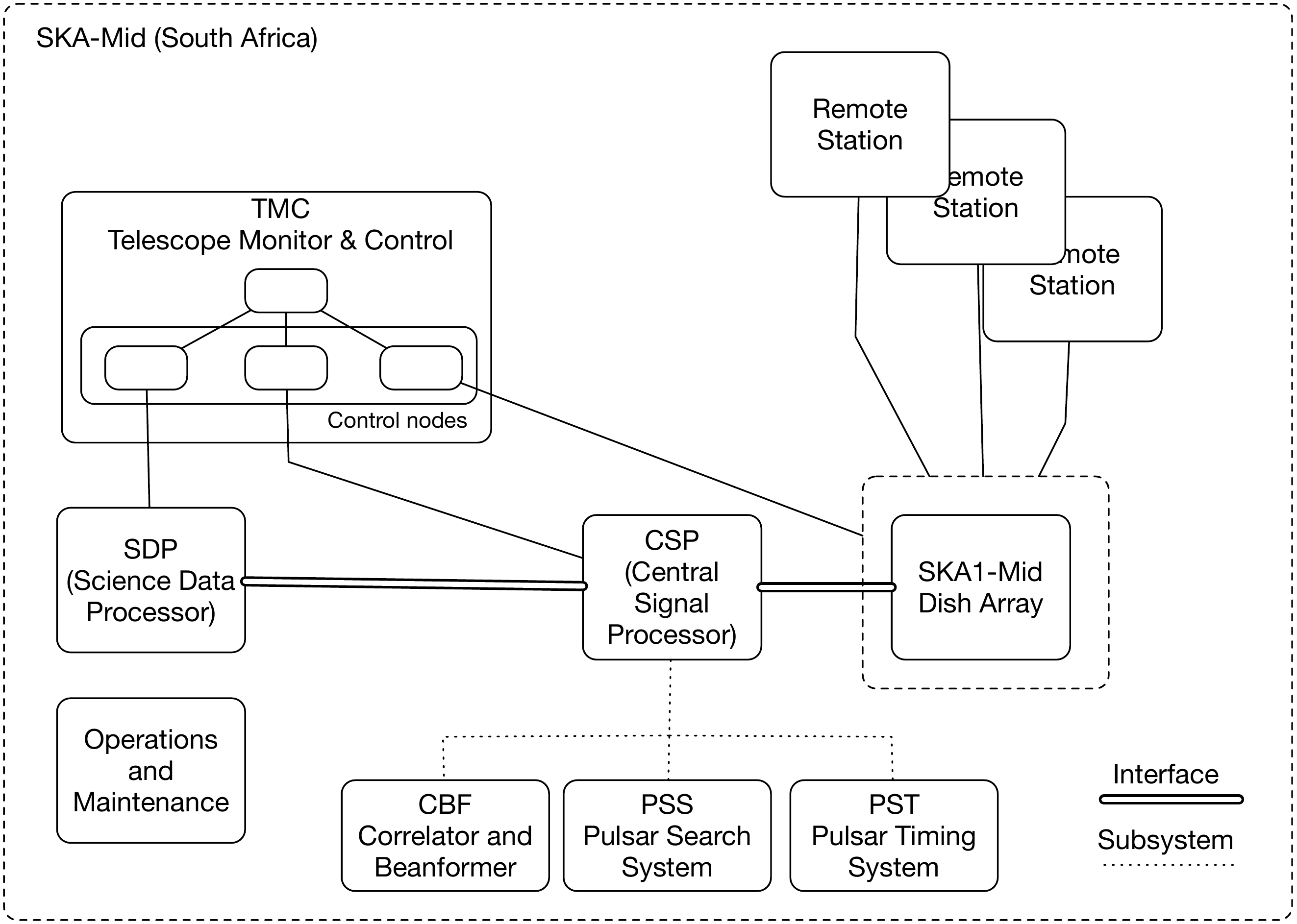}
	\caption{Simplified model of the SKA system architecture, including the key components for its operation. Only the SKA-Mid subsystem is presented here. The SKA-Low radio telescope subsystem architecture is similar to it but with a low-frequency aperture array (arrays of dipole antennas) instead of dish arrays.}
	\label{fig:ska-architecture}
\end{figure}

The architecture of the SKA involves two large-scale subsystems, namely the Mid and Low radio-telescopes, which capture low and mid frequencies in the electromagnetic spectrum through arrays of dipole antennas and arrays of dish observatories. As described in Figure~\ref{fig:ska-architecture}, both subsystems have equivalent architectures, where the raw data is captured through the respective arrays and turned into \textit{visibilities} (pulsar survey candidates and their timing) through the \textit{Central Signal Processor} (CSP). The vast amount of data generated by the CSP is ingested, in turn, by the \textit{Science Data Processor} (SDP), where data is reduced and packed so scientists can make decisions in near real-time, e.g., about noise in the captured data.

\subsubsection{SKA architecting and development process}\label{sec:dev-process}

During its pre-construction phase, the system was partitioned in multiple subsystems and then delegated to eleven international consortia for their detailed design, using ICDs to have a clear separation of responsibilities.  In the SE process followed in this phase, the compliance of the design against the Subsystem Level (L2) requirements (derived from the System Level (L1) ones in this phase) and the ICDs were rigorously supervised. This phase also involved a number of preliminary design reviews of the architecture and was closed with a formal Critical Design Review (CDR), where the final design of each subsystem is presented through completed analysis, simulations, schematics, software and test results~\cite{fowler2014developing}. 

The transition between the CDRs and the construction phase, which started at late 2019 and is coming to a closure as this paper is written, is described as the \textit{bridging} phase. Interestingly, during this bridging phase, the organization opted to adopt the \textit{Scaled Agile Framework} (SAFe)~\cite{leffingwell2018safe} approach to tackle the challenge of developing the software-intensive elements aligned to a common vision, after the pre-construction consortia got dissolved. This involved, in general (1) forming an \textit{Agile Release Train} (ART), that is to say, cross-disciplinary agile teams that will work towards such a common vision, (2) defining and prioritizing a backlog of features for each \textit{Program Increment} (PI), which are development sprints planned by the aforementioned teams during the \textit{PI-planning} meetings, and (3) driving each team work following a \textit{Scrum} process. The SKA is a unique case, among systems of this scale, where traditional SE practices (particularly the ones followed in the pre-construction phase), meet with the agile principles and the practices described above. This means, for instance, that the requirements gathered during the consortia phase (both L1 and L2) were used as an input for defining the \textit{Solution Intent (SI)}, an evolving knowledge base that serves as the `single source of truth' about the product vision, what is being built, and how it will be built. This transition from SE to an agile process also implies the need for managing two types of interfaces: the high-level ones originally defined through the ICDs, and the low-level ones created within the cross-disciplinary teams, and specified within the Solution Intent. %

\section{Study Design}\label{sec:study-design}

\subsection{Study goals and research questions}

The goal of this study, as a follow-up of the exploratory case study on LOFAR+, is to \textit{confirm and extend the practices that harmonize architecting processes of Systems Engineering (SE) and Software Engineering (SWE) in large-scale SoS}. A more precise definition of this goal, using the five parameters proposed by the Goal-Question-Metric (GQM) goal template~\cite{basili1992software} is as follows:

\begin{quote}
	\textbf{Analyze} \textit{the architecting process of an ultra-large-scale SoS involving multiple organizations} 
	\textbf{for the purpose} \textit{of confirming the practices for harmonizing SE and SWE architecting processes identified in the context of the LOFAR+ case study, and identifying new ones}
	\textbf{with respect to} \textit{cross-disciplinary issues related to requirements, subsystem design decisions, and subsystem interfacing}
	\textbf{from the viewpoint of} \textit{engineers}, %
	\textbf{in the context of} \textit{the Square Kilometre Array (SKA) project}.
\end{quote}

Based on this goal, we derived the following research questions:

\begin{mdframed}[innertopmargin=-2pt]
\paragraph{\textbf{RQ1}}  \textit{To what extent have the harmonization practices identified in LOFAR+, successfully addressed the same kind of issues in the context of the SKA, on requirements (RQ1.1), subsystem design decisions (RQ1.2), and subsystem interfacing (RQ1.3)?}
\paragraph{\textbf{RQ2}}  \textit{What additional cross-disciplinary harmonization practices, related to requirements (RQ2.1), subsystem design decisions (RQ2.2) and subsystem interfacing (RQ2.3), were adopted in SKA?  What additional practices were adopted to address the issues that remained un-solved in LOFAR+ (RQ2.4)?} 
\end{mdframed}

Along the same lines of the exploratory case study~\cite{cadavid2021casestudy}, whose results this study aims to confirm and extend, these research questions entail investigating them within its real-life context. Consequently, we opt for a \textit{confirmatory case study} as an empirical research method~\cite{runeson2009guidelines}. 

\subsection{Data Collection}\label{sec:datacollection}

The SKA, as described in the previous section, is a massive project that involves hundreds of engineers around the world. %
To better understand the case subject (i.e.~the SKA project) and to define properly tailored research instruments, we conducted an \textit{in-situ} pre-study by analyzing relevant archival data of the project, e.g., the current (at the time) Solution Intent document, and most importantly, by participating as observers on the 10th PI-planning event (taking place virtually due to the COVID-19 pandemic). 
The former allowed us to gain an up-to-date understanding of the principles followed by the SKA for its evolving design and implementation. The latter helped us gather details (also related to such principles) about the early stages of the SKA, and the way requirements, design decisions and interfaces were managed in the project. The event's onboarding sessions, where questions were proposed and discussed as relevant topics arose, were particularly useful for this purpose. Separate meetings with the SAFe consultant of the project also served as a source of information about the transition of the project into this framework.

Based on the findings of this pre-study, we designed a \textit{questionnaire-based survey} for data collection purposes.
The survey instrument was tailored to the target participants, that is to say, considering the specifics of the SKA context and the terminology used by the engineers involved in the Agile Release Train. 
As described in Table~\ref{tab:mapping}, the survey questions were organized in four sections: \textit{demographic data}, \textit{requirements-related issues and practices} --- \practiceref{B1}, \textit{design decisions-related issues and practices} --- \practiceref{B2} and \practiceref{B3}, and \textit{interface-related issues and practices} --- \practiceref{B4}.
Questions concerning the handling of issues identified in LOFAR+ without best practices to address them, i.e.~\problemref{P8} and \problemref{P9}, were added to the survey sections relevant to them, namely the second and fourth sections of the survey.

We used a combination of closed and open-ended questions in order to achieve in-depth insights. Most of the closed questions used a five-point Likert scale to measure perceptions pertaining to, for example, the impact of a given practice in the context of the SKA. Given the broad range of elements addressed by the investigated practices (requirements, design decisions, interfacing), and the low likelihood as we perceived it of all respondents to be familiar with all of them (especially due to the system scale), an explicit `Do Not Know' option was added as a sixth option in the answer scales. This allowed to avoid mixing \textit{no opinion} responses with \textit{neutral/low-score} ones.

\begin{table}[t]
	\centering
    \footnotesize
    \renewcommand{\arraystretch}{1.2}
	\begin{tabularx}{\linewidth}{cXlc}
		\toprule
		\bfseries RQ & \bfseries Scope &  \bfseries Survey questions & \bfseries \begin{minipage}[c]{0.14\textwidth}\centering LOFAR+ Practices\end{minipage}\\
		\midrule
		N/A & Demographics & Q1.1-Q1.7 & \\ \hline%
		RQ1.1 & Requirements  & Q2.1-Q2.4, Q2.6 & \practiceref{B1}\\%
		RQ2.1 & Requirements & Q2.7 & \\ \hline
		RQ1.2 & Design decisions  & Q3.1-Q3.5 & \practiceref{B2},\practiceref{B3}\\%
		RQ2.2 & Design decisions & Q3.6, Q3.7 & \\ \hline
		RQ1.3 & Interfacing  & Q4.1, Q4.2-Q4.5 & \practiceref{B4} \\%
		RQ2.3 & Interfacing & Q4.5 & \\ \hline
		RQ2.4 & Requirements and Interfacing (issues not addressed in the LOFAR+ study) & Q2.5, Q4.3, Q4.6, Q4.7 & 	\\	
		
		\bottomrule
    \end{tabularx}

	\caption{Mapping between research and survey questions, including the issues and best practices explored per question.}
	\label{tab:mapping}
\end{table}

The survey was piloted by three ASTRON engineers that also participated on the LOFAR+ case study, and were familiar with the SKA project by being partially involved on it. These three participants were not included in the sample population of the survey, and neither was any of the other LOFAR+ study participants. With the input received from the pilot's participants, the structure, wording and the use of the terminology were improved for clarity. The survey was later reviewed by one of the SKA's lead software architects, who also contributed to its improvement. Furthermore, the survey activity was validated by the SKA Organization's head of Mission Assurance. The final version of the survey was developed and published online through the Qualtrics.XM platform\footnote{https://www.qualtrics.com/}. The target population, which consisted of SKA engineers that work under the SAFe framework, was profiled with the help of the head software architect, who estimated it in around 140 people. He also customized a mailing list to match the target population and used it to submit invitations to participate. This was followed by a number of invitations through the SKA's Slack group, and through personal interventions in activities within the 10th PI, as discussed above.

The final form of the
questionnaire is available in the study replication package\footnote{\url{https://figshare.com/s/4683df26c4283f9e32e1}\label{fn:replication}}.

\subsection{Data Analysis}
\label{sec:dataanalysis}

The data collected from the survey were analyzed using descriptive statistics for the closed questions, and Qualitative Content Analysis (QCA) for the open-ended ones. In the latter, an inductive approach was followed~\cite{elo2008qualitative}, meaning that the identified categories and groups of codes emerged from a systematic process of open-coding, categories creation, and abstraction. The most prominent categories together with representative quotes can be found in Appendix~\ref{sec:appendix}, and the full data in the replication package of this study (see Footnote~\ref{fn:replication}). The results of the data analysis are discussed in the following section. 

\section{Results}\label{sec:results}
\begin{figure}
	\centering
	\includegraphics[width=.9\linewidth]{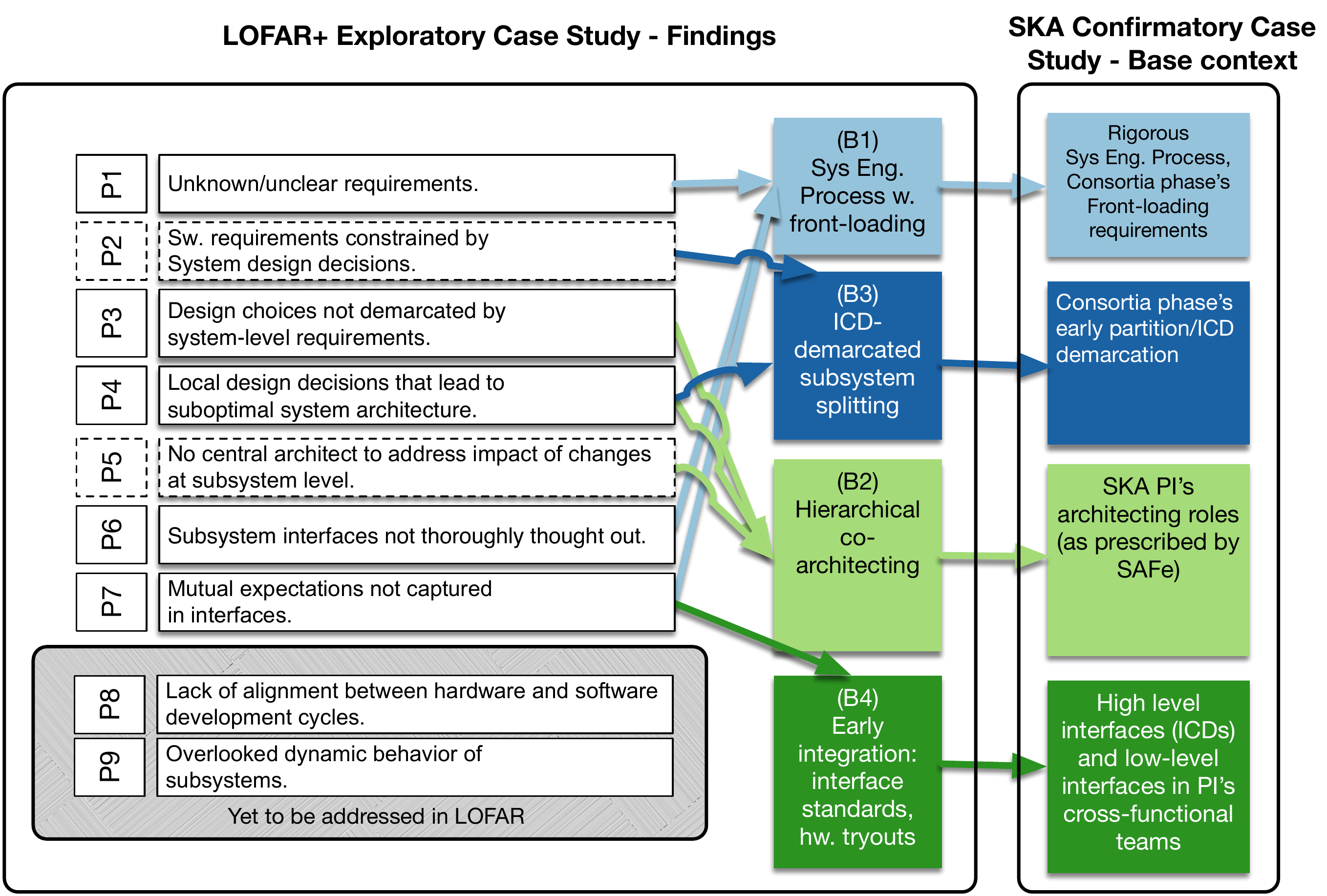}
	\caption{Summary of the SKA practices that reflect the ones identified in the LOFAR+ case, as identified by the pre-study analysis results.}
	\label{fig:basecontext}
\end{figure}

In the following, we present our analysis of the survey responses, which will be subsequently used in Section~\ref{sec:discussion} to answer the proposed research questions. To this effect, the presentation follows the structure of the survey, i.e.~discussing demographic data, requirements-related issues and practices, design decisions-related issues and practices, interface-related issues and practices, and practices for the issues identified as not solved in the LOFAR+ case. With the exception of the subsections that describe the demographics and the practices for unsolved issues on LOFAR+, each of the other subsections are introduced with a description of the specific practices within the SKA that reflect the ones defined in the LOFAR+ case. This mapping between SKA and LOFAR+ practices, which resulted from the pre-study conducted based on Observational and Archival data as described in the previous section, is summarized in Figure~\ref{fig:basecontext}.

\subsection{Demographics}\label{sec:demographics}

\begin{figure}[t]
	\centering
	\includegraphics[width=\linewidth]{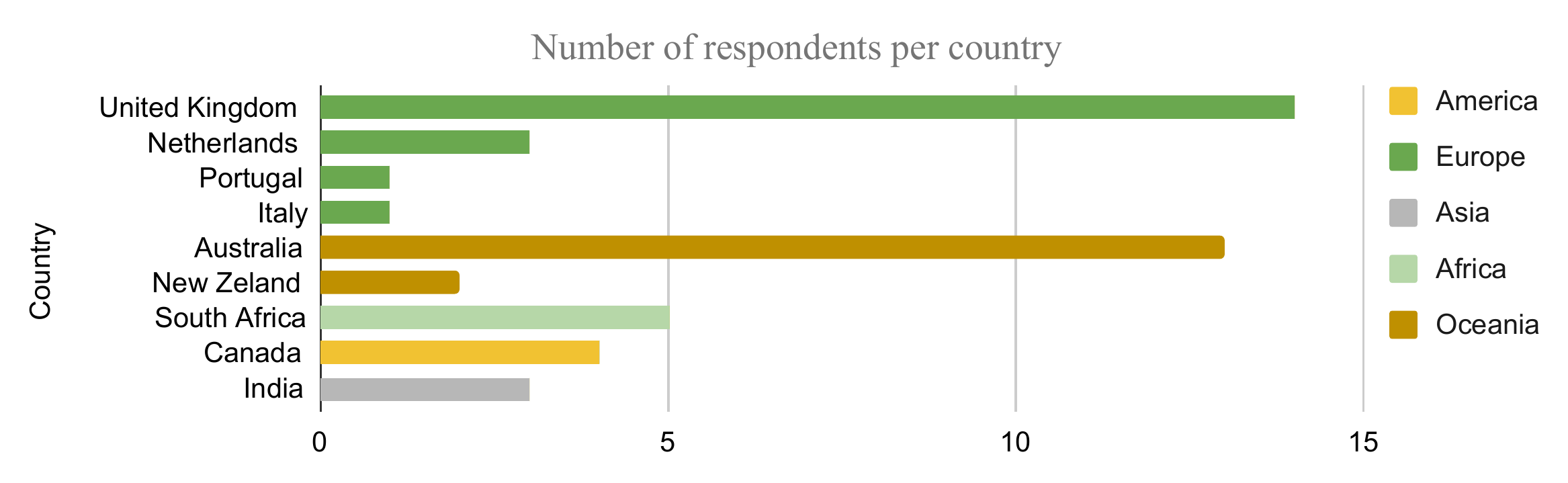}
	\caption{Distribution of the respondent's organizations across the world.}
	\label{fig:orgs-world-dist}
\end{figure}

\begin{figure}[t]
	\centering
	\includegraphics[width=.9\linewidth]{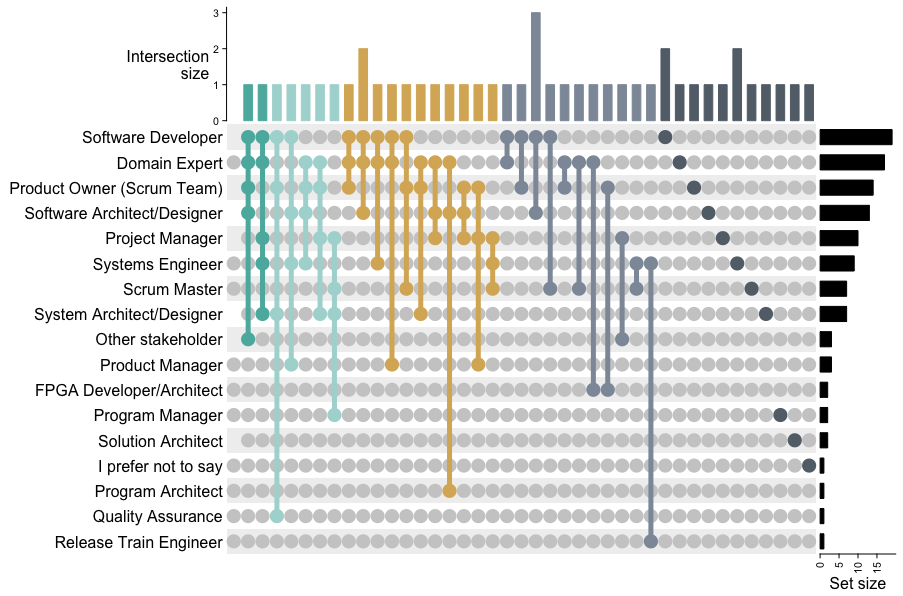}
	\caption{Roles performed in the SKA by the survey respondents}
	\label{fig:14roles-UpSet}
\end{figure}

A total of 46 responses were collected between late April and early September of 2021, out of the total target population of around 140 as discussed in Section~\ref{sec:study-design}. The base organizations of the respondents are distributed across the globe as illustrated in Figure~\ref{fig:orgs-world-dist}. The respondents have an average 18.13 years of work experience, with 5.6 years of experience in the SKA project. As seen in Figure~\ref{fig:14roles-UpSet}, most of the respondents identify themselves with multiple roles within the project, with \textit{Software Developer} (20\%), \textit{Domain Expert} (15\%), \textit{Product Owner} (14\%) and \textit{Software Architect} (12\%) being the most prominent ones.


The information collected by the demographics section of the survey provides insights on the opted-out (i.e. `Do Not Know') responses to closed questions (Likert scale) in the rest of the survey; on average 12.65 participants opted out per closed question. In particular, by running Spearmann correlation tests between the respondents' number of opted-out responses and their years of overall and SKA-specific experience, we found only a slight negative correlation for the SKA experience (rho $=-0.32$ with a p-Value of $0.032$) and weak evidence for the overall experience (rho $=-0.14$ with a p-Value of $0.35$). Therefore, in general, experience levels do not sufficiently explain the number of opted-out responses presented in the following sections. However, by grouping the roles in more general ones\footnote{\textbf{Architects}: Solution Architect, Program Architect, System Architect/Designer, Software Architect/Designer; \textbf{Engineers/Developers:} Systems Engineer, Software Developer, Scrum Master, Quality Assurance, Product Owner; \textbf{Managers}: Project Manager, Program Manager, Release Train Engineer; \textbf{Others:} Other roles, Other Stakeholder, \textit{I prefer not to say}, Domain Expert.}, namely \textit{Architects}, \textit{Engineers/Developers}, \textit{Managers}, and \textit{Others}, and counting the opted-out questions by each group, we found that most opt-outs came from two groups: managers, and engineers/developers as summarized in Table~\ref{tab:opted-out-demog}. Arguably, this could be attributed to two different phenomena: for the managers, due to the lack of sufficient depth of technical details that some of the questions required; for the engineers/developers, due to the very specific roles within the project performed by this part of the population allowing them to focus only on certain parts of the survey. Both explanations are consistent with our decision of including an explicit `Do not know' option on the Likert scales, as described in Section~\ref{sec:datacollection}.

\begin{table}
	\centering
	\footnotesize
	\renewcommand{\arraystretch}{1.2}
	\begin{tabularx}{0.8\textwidth}{Xcccc}
		\toprule
		\bfseries Role group &  \bfseries Group size  &  \bfseries Total opt-outs &  $\mathbf{\bar{x}}$ & $\mathbf{\sigma}$   \\ \midrule		
		Managers & 12 & 93 &  0.28  & 0.12  \\ 
		Engineers/developers & 35 & 269   & 0.27 & 0.07 \\ 
		Architects & 20 &  122 & 0.22 & 0.09  \\			
		Other roles & 23 & 133  & 0.21  & 0.11 \\ 
		 \bottomrule					
	\end{tabularx}
	\caption{Mean and standard deviation of the number of `Do Not Know' answers selected across the 28 Likert questions of the survey by groups of roles, normalized by the corresponding groups size.}
	\label{tab:opted-out-demog}
\end{table}

\subsection{Practices and issues related to requirements}

In the LOFAR+ case, the issues of \textit{unclear requirements} \problemref{P1},  \textit{interfaces that are not thoroughly thought out} \problemref{P6} or that \textit{do not capture the mutual expectations of their related parties} \problemref{P7} were addressed by adopting a \textit{more rigorous systems engineering process, with front-loaded requirements} \practiceref{B1}. As depicted in Figure~\ref{fig:basecontext}, this practice can be seen as already followed in the SKA, as the project defined and allocated up-front, in the consortia phase described in Section~\ref{sec:SKA-overview}, a series of System Requirements (known as L1 requirements) and Element (or sub-system) requirements (known as L2 requirements)~ \cite{santander2017agile}. However, for the SKA construction, part of the software-intensive elements of the SKA are being developed as a Solution Intent, following SAFe, which encourages leaving room for an emerging understanding of specific requirements, based on intent. Given this, we explore practice \practiceref{B1}, in the context of the SKA, by investigating how these front-loaded  requirements (i.e., L1 and L2) helped or hindered the Solution Intent. Furthermore, we investigate what additional practices have been adopted by the SKA to address requirements related issues, including \problemref{P1}, \problemref{P6} and \problemref{P7}.

\begin{figure}[t]
	\centering
	\includegraphics[width=.75\linewidth]{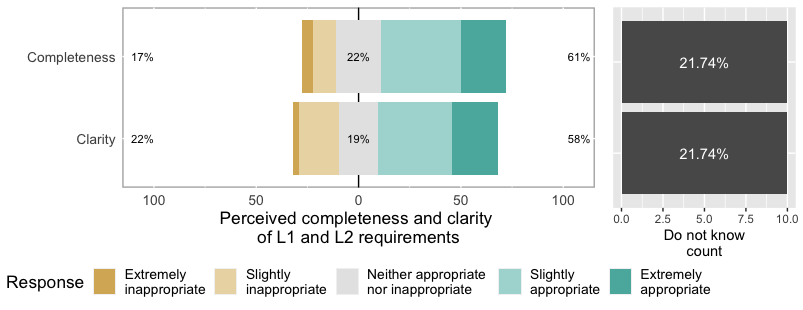}
	\caption{Distribution of the responses to Q2.1 about the perception of clarity/completeness of L1/L2 requirements in the SKA project, using as a diverging point the neutral opinion. The opt-out selection (shaded) does not belong to the scale but is included as a reference.}
	\label{fig:21likert-clarity-comp}
\end{figure}

\begin{figure}[t]
	\centering
	\includegraphics[width=.75\linewidth]{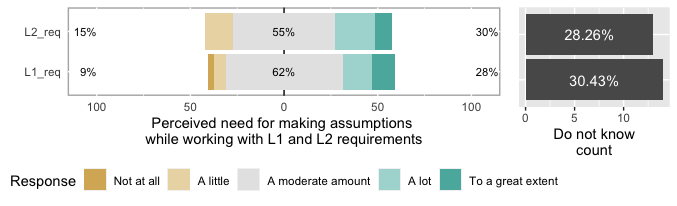}
	\caption{Distribution of the responses to Q2.3 about the extent to what the participants and their teams had to make assumptions with the requirements elicited from the Systems Engineering process.}
	\label{fig:23l1l2assumptions}
\end{figure}

\begin{figure}[t]
	\centering
	\includegraphics[width=.75\linewidth]{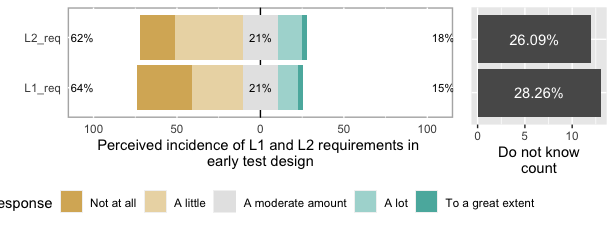}
	\caption{Distribution of the responses of Q2.2 about the extent to what L1 and L2 requirements have been used for early tests design.}
	\label{fig:22reqsntesting}
\end{figure}

According to the results shown in Figure~\ref{fig:21likert-clarity-comp}, and using a neutral opinion as the diverging point, most of the respondents (excluding the ones that opted out of the question)  evaluated the completeness and clarity of L1 and L2 requirements \problemref{P1} as appropriate or extremely appropriate (61\% and 58\% respectively). However, when looking at the degree to which participants have had to make assumptions while working with these requirements (see Figure~\ref{fig:23l1l2assumptions}), and to which these were used in the early tests design (see Figure~\ref{fig:22reqsntesting}) 
as two of the consequences of having incomplete/unclear requirements according to the LOFAR+ case, 
this seems rather counter-intuitive. As seen in Figure~\ref{fig:22reqsntesting}, despite the perceived clarity and completeness of the requirements, around 85\% of the respondents (excluding the ones that opted out of the question) feel that L1/L2 requirements had a moderate to no influence on the design of tests early in the development process. Furthermore, as described in Figure~\ref{fig:23l1l2assumptions}, most of the respondents perceive that working with this kind of requirements involves making assumptions to a moderate extent, especially regarding system-level (L1) requirements (62\%). Given the positive perception of the clarity and completeness of the requirements defined up-front \practiceref{B1}, the results regarding the need for making assumptions when working with system-level requirements, and their influence of early testing, were not expected. We therefore look into the open-ended questions for a more nuanced analysis of the practitioners' perceptions.


The analysis of the 38 open-ended responses to Question 2.6, about the perception of the positive or negative impact of these front-loaded requirements \practiceref{B1} in the Solution Intent, shows a predominant positive view of them as helpful, albeit to a different extent (codes with a total groundedness\footnote{The groundedness $\groundedsym$ of a code refers to the number of quotations extracted during the coding phase that ended being associated with it through the inductive Qualitative Content Analysis process (from which such a code emerged)~\cite{friese2019qualitative}.} $\groundedsym$ of 36). There is also a smaller minority with a negative view due to a perceived lack of actual influence of said requirements in the Solution Intent (codes with $\groundedsym=5$). Both points of view are described in the following.

\subsubsection{Views that confirm the practice}

According to the quotes assigned to the emergent categories as \coderef{Helpful in General}{10}, \coderef{Solid Ground}{7} and \coderef{Solution Intent Base}{4} as described in Table A.1 (in the online Appendix --- see Section~\ref{sec:dataanalysis}), L1/L2 requirements are key for ultra-large-scale, long-term projects like the SKA as they provide an initial partitioning criteria, making them more manageable \problemref{P6}. Furthermore, the requirements provide vital elements that should flow down to the system design and to the Solution Intent; in particular these elements are edge cases, data formats, and `nice-to-have' features \problemref{P1}. More importantly, though, the requirements provide information in advance regarding formal testing \problemref{P1}, although this does not seem to be always the case as shown in Figure~\ref{fig:22reqsntesting}.

The responses coded as \coderef{Transition to Agile Issues}{5}, on the other hand, highlight that L1 and L2 requirements while helpful are also in danger of being inconsistent as the Solution Intent evolves, and (eventually) they will clash with reality once the instrument is built. For this reason, there is a perceived danger of losing the focus during the agile process as the requirements are not serving as the framework to govern the development in the long term. Respondents also highlight the need to fine-tune this transition from front-loaded requirements to Program Increment: they mention that, in retrospect, moving from requirements-driven to an agile development approach made the project slow down, and that L1 requirements could have been defined (if such an approach switch were foreseen) in a way that suits better the adopted agile framework.

Similarly, the responses with the \coderef{Missing Intermediate Elements}{6} code pointed out the need for having additional intermediate elements between the front-loaded requirements and the agile process. It is perceived that the project requires better traceability between L1/L2/L3 requirements (where L3 requirements refer to implementation specifics), and the features/tasks the agile teams are working on. According to the respondents, this would be useful to identify if the understanding of the system ---the one that emerges from the agile process--- is deviating from the original system definition. At the same time, and given the very long term of the project, it is perceived that not having requirements for intermediate releases (in other words, short- and medium-term requirements) hinders the Solution Intent. According to the respondents, \textit{requirements for the intermediate versions of the system are key} to properly drive (accordingly to the final solution vision) the architecture, and for having proper test specifications of the software systems. Furthermore, it is perceived that the long-term requirements do not always capture the activities needed for specific intermediate phases such as construction and commissioning.

Finally, the code \coderef{Imperfect Req Issues}{3} groups the comments where respondents pointed out that L1 and L2 requirements, when not well defined, result in constant refactoring. Some respondents describe this as requirements with a lot of elements left to be assumed or discovered. It was pointed out that, for a proper requirements tree, an iterative validation process of the flown-down requirements completeness, i.e., from L1 down to L2 and L3, and all the way back up, would be necessary.

\subsubsection{Views that do not confirm the practice}

According to the respondents whose responses were coded as \coderef{Little Weight}{2} or \coderef{Not Followed}{3}, as described in Table A.2, L1/L2 requirements have little weight or are ignored during decision making in particular cases. %
Respondents mentioned two perceived causes for this: this kind of requirements are not useful for a lean/agile process, or in some cases teams do not follow them due to budget and time constraints. Other respondents mentioned ---arguably as a consequence of the above--- that the Solution Intent (at the moment of filling in the survey) was driven more by assumptions and experience of the architects rather than the L1/L2 requirements.

\subsubsection{Additional practices adopted by SKA to address requirements-related issues}

According to the analysis of the 31 open-ended responses registered for Question 2.7 about how requirements-related issues like unclear and incomplete requirements are being prevented or mitigated in the SKA, the practices followed in this direction are mostly mitigation-oriented. A number of respondents, as seen in code \coderef{Agile Filling Gaps}{9} (see Table A.4), pointed to SAFe as the key strategy to mitigate incomplete or unclear requirements given the realization that L1 and L2 requirements inevitably evolve over time, and that in an agile framework like SAFe, evolving requirements are expected to be part of the basic understanding of the system. In the same question, other respondents (code \coderef{Prototypes Testing}{5}) see test-driven design as a key agile practice to identify requirement issues as the construction progress. This has been perceived at two different stages: early during the definition of the acceptance tests, and while receiving feedback from stakeholders (often involved in the definition of the original requirements) through prototype testing and demos. A few respondents, on the other hand, highlighted that the SKA already mitigates these issues with significant efforts on requirement reviews, where traceability, consistency and completeness are evaluated, and assumptions documented  (\coderef{Reqs Reviews}{4}). 

\subsection{Practices and issues related to subsystem design decisions}\label{sec:design-related-practices}

In LOFAR+, \textit{local design decisions had a negative impact on the overall system qualities and architecture} \problemref{P4}, as such decisions were \textit{not properly demarcated} \problemref{P3}, and due to the \textit{lack of a central architect role} \problemref{P5}. Proper subsystems splitting and demarcation with ICDs \practiceref{B3} and having hierarchical co-architecting roles \practiceref{B2} were two practices identified as key to address these issues.

These two practices can be seen as already applied on the SKA, as (1) during its consortia phase (see Section~\ref{sec:SKA-overview}) the system was subject to such a partition/ICD demarcation, with the resulting architecture being the baseline of the project's Solution Intent, and (2) the adopted SAFe framework prescribes such hierarchical co-architecting roles (solution/system architects). Given this, as also depicted in Figure~\ref{fig:basecontext}, we explore practices \practiceref{B3} and \practiceref{B2} in the context of the SKA by investigating the perceived impact of this early partition, and the hierarchical architecting roles on the Solution Intent, on the system architecture and qualities. Furthermore, we investigate further practices adopted by the SKA to address design decisions-related issues, including the use of non-functional requirements (NFR) as the means prescribed by the SAFe framework to properly demarcate the design decisions \problemref{P3}.

\begin{figure}[t]
	\centering
	\includegraphics[width=.75\linewidth]{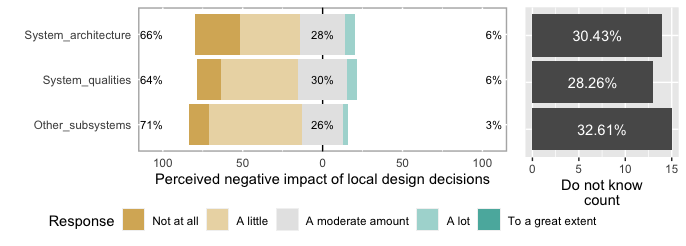}
	\caption{Distribution of the responses of Q3.1 about the extent to what local design decisions have had a negative impact on other subsystems, system-level qualities or the overall system architecture.}
	\label{fig:31local-des-dec-impact}
\end{figure}

\begin{figure}[t]
	\centering
	\includegraphics[width=.75\linewidth]{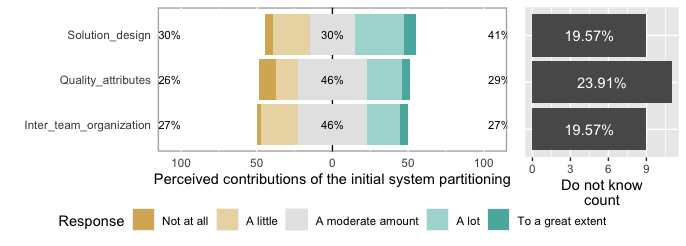}
	\caption{Distribution of the responses of Q3.4 about the extent to what the early partition of the system has contributed to the overall solution design, the system qualities and the inter-team organization.}
	\label{fig:34initialpartcontribution}
\end{figure}

\begin{figure}[t]
	\centering
	\includegraphics[width=.75\linewidth]{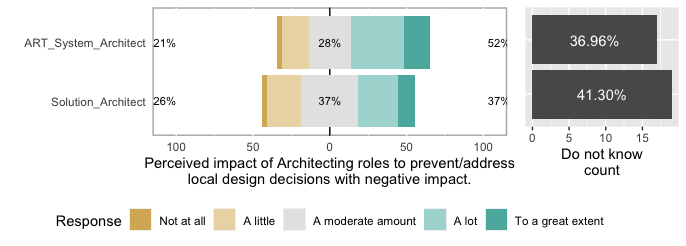}
	\caption{Distribution of the responses of Q3.5 about the extent to what the hierarchical architecting roles prescribed by SAFe contribute to prevent or address the negative impact of local design decisions on the overall system qualities.}
	\label{fig:35archrolespreventndd}
\end{figure}

As it can be seen in Figure~\ref{fig:31local-des-dec-impact}, the negative impact of the local design decisions on the system design and its qualities is perceived, for the most part, as small to moderate. This suggests that the SKA is seemingly doing a good job when it comes to this particular issue previously identified in the context of LOFAR+ \problemref{P4}. However, given the perceived contribution of the early partition of the system \practiceref{B3}, and the hierarchical architecting roles \practiceref{B2} to mitigate issues related to local design decisions (described in Figures~\ref{fig:34initialpartcontribution} and~\ref{fig:35archrolespreventndd} respectively), the latter seemed to have a more significant influence in the positive perceptions on that respect.

As a result of the analysis of the 28 open-ended responses recorded for Question 3.2, two main categories of design decisions (Table A.5), were perceived as having a negative impact on the overall system design and its qualities%
. The first and one of the most dense ones, \coderef{Prescriptive Decisions}{8}, corresponds to a number of \textit{technical, often prescriptive cross-cutting design decisions} with a perceived negative impact on the system due to the constraints or limitations they posed when implementing the related subsystems. Table~\ref{tab:2-5-presc-on-dd} lists a summary of specific design decisions of the aforementioned category, as described by the survey  participants (the referenced subsystems are described in more detail in Section~\ref{sec:SKA-overview}). It is worth highlighting that some of these design decisions, which contributed to an overly-complicated architecture, are linked to the initial partition of the SKA ---related to Practice \practiceref{B3}--- as its root cause, e.g., the seemingly unnecessary abstraction layers.

\begin{table}
	\centering
	\footnotesize
	\renewcommand{\arraystretch}{1.2}
	\begin{tabularx}{\textwidth}{XX}
		\toprule
		\bfseries Design decision &  \bfseries Consequence                \\ \midrule		
		Algorithmic prescriptions on the signal processing pipeline (e.g., filter coefficients)    & Global signal processing optimization alternatives limited. \\ \hline
		Delegating computation responsibilities at a certain system level (e.g., delay calculations) &  Workarounds to overcome the limitations of the selected system level.                          \\ \hline
		Data (visibilities) resolution &  Mismatch between the expected resolution and the limitations of the devices                \\ \hline
		Transport Protocol selection (e.g., between the CSP and the SDP) &     Interfacing/Performance issues                       \\ \hline
		Additional abstraction layers for integration (e.g., for the TCP) &  Over-complicated design (e.g., unnecessary pass-through nodes in the system )                 \\ \hline
		Use of common base-classes across sub-systems &  Broken deployments and large maintenance efforts when such classes are changed.                  \\ \hline			
		Deployment infrastructure selection (e.g., orchestrated containers for both TMC and and SDP) &    Compatibility/performance issues                      \\  \bottomrule

	\end{tabularx}
	\caption{Summary of the `prescriptive', cross-cutting design decisions that had a perceived negative impact on the system design and its qualities. Subsystem names refer to Figure~\ref{fig:ska-architecture}.}
	\label{tab:2-5-presc-on-dd}
\end{table}

 The quotes of the second most prominent code (\coderef{Local Interpretations}{8}), pointed out that some decisions, when based on the team member's previous knowledge on what makes more sense to them, or on their own interpretation of the requirements, have had an impact on qualities like availability, maintainability, and flexibility. In addition, this kind of decisions often leads to extra effort during integration, e.g., when integrating subsystem components that are originally operating on their own local networks into a single, large network. From the responses related to this code, it is worth noting that: (1) these are closer to the issues identified in the LOFAR+ case \problemref{P4}, where design decisions made independently at the subsystem level have an impact on the system architecture and its qualities; and (2) they contradict, to some extent, the point of view of the comments of the previous category, where some early, prescriptive decisions were an issue, as they highlighted that having a holistic view of the system defined early on would have helped to make better design decisions.

\subsubsection*{Additional practices adopted by the SKA to address subsystem design decisions-related issues}

Most of the quotes extracted from the 26 open-ended responses to Question~3.6 (Table A.6) about the practices adopted by the SKA to prevent local design decisions having a negative impact on the overall system architecture/qualities \problemref{P4}, were classified as communication-oriented ones (\coderef{Communi-cation Related}{9}). It needs to be noted however that such practices are still seen as a work in progress. The relevant quotes agree on the benefits of enabling higher levels of transparency within the project through \textit{active communication across levels} (e.g., to get input regarding potential impact of the design decisions), and by \textit{facilitating access to key roles and groups}, namely systems engineers, solution architects, architecture groups and communities of practice. From this series of responses, it is worth highlighting that some of these communication-oriented practices emerged as a measure for the perceived risks of the `siloed' architecture: local design decisions that could be made without considering the impact in the other subsystems \problemref{P4}. This, on the other hand, is consistent with the perception of the respondents about the influence of the hierarchical architecture \practiceref{B2} within the SAFe process as previously discussed: according to the respondents, this risk has been mitigated with architecting roles, e.g., working towards consensus-driven common solutions by engaging the architectural experts on both sides of the ICDs.

\subsection{Practices and issues related to subsystem interfacing}\label{sec:interfacing-pract-prob}

In the LOFAR+ case study, the interfacing between subsystems was identified as a significant source of issues for the interplay between different engineering disciplines, and in particular on how their documentations were not capturing the mutual expectations of the involved parties \problemref{P7}. To address these issues, early integration practices \practiceref{B4} were identified. In this case, the pre-study did not provide enough information on which specific early integration practices have been followed in the SKA. Therefore, to explore \practiceref{B4} in the context of the SKA, we investigated to what extent the interfaces within the SKA are subject to \problemref{P7}, and which particular practices have been followed to address such an issue. We consider both high-level and low-level interfaces in the project (as depicted in Figure~\ref{fig:basecontext}):  the ICDs defined during the project's consortia phase, and the interfaces used within the ART's cross-functional agile teams. Furthermore, for the former, we take into account that in the SKA, ICDs were refined  through a number of formal design reviews, as prescribed by the systems engineering practices~\cite{fairley2019systems}.

\begin{figure}[h]
	\centering
	\includegraphics[width=.75\linewidth]{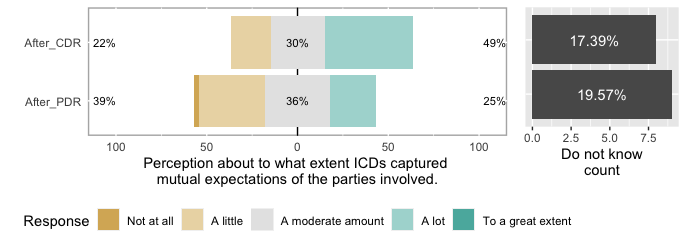}
	\caption{Distribution of the responses of Q4.1 about the extent to what interfaces defined in the early partition of the system captured the mutual expectations of the parties involved.}
	\label{fig:41mutual-exp-dact}
\end{figure}

\begin{figure}[h]
	\centering
	\includegraphics[width=.75\linewidth]{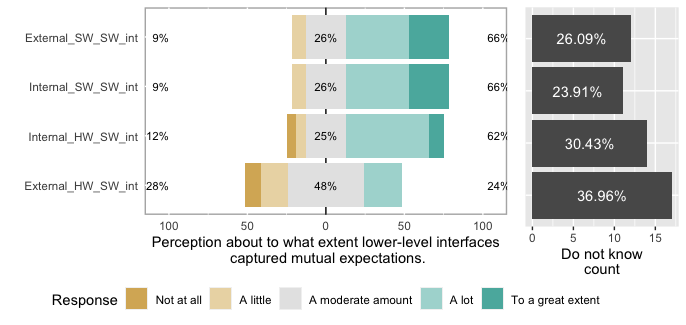}
	\caption{Distribution of the responses of Q4.4 about the extent to what lower-level interfaces, within and between cross-functional teams captured the mutual expectations of the parties involved.}
	\label{fig:44internalintmexp}
\end{figure}

The perception of the extent to which the ICDs in SKA captured the mutual expectations of the involved parties, captured in Question 4.1 and illustrated in Figure~\ref{fig:41mutual-exp-dact}, suggests two things. First, the \textit{Critical Design Review} had a seemingly significant impact on this particular aspect of the ICDs, and second, the overall perception of the ICDs ---regarding how well they captured mutual expectations after said formal reviews--- is mostly positive. Regarding the lower-level, internal interfaces, as seen in Figure~\ref{fig:44internalintmexp}, software-to-software interfaces either within the same agile team, or between separate ones, have a highly positive perception when it comes to the way they capture mutual expectations. For hardware/software interfaces, the perception (with exception of internal ones) is significantly more negative (mostly little to moderate capture of mutual expectations) regarding interfaces used by separate teams.

Most of the codes extracted from the 29 responses given to Question 4.2, about \textit{what was missing on the ICDs to properly capture the mutual expectations of the involved parties}, are related to either missing elements on the ICDs (multiple codes with a total groundedness of 14), or the ICDs' life-cycle (with a single code of \groundedsym=7), as listed in Table A.7. In the former, four quotes (\coderef{Incomplete Cases}{4}) mentioned specific examples of incomplete or vague ICDs within the SKA (e.g., TM to LFAA\footnote{Telescope Manager to Low Frequency Aperture Array}, TMC\footnote{Telescope Management and Control} to Elements, CSP's LMC\footnote{Central Signal Processor's Local Monitor \& Control} to sub-elements --- see Figure~\ref{fig:ska-architecture}) without going into specifics. The other ten quotes (coded as \coderef{Missing Elements}{10}) pointed to the need of further context on how the subsystems related to an ICD would operate within the overall system, and an end-to-end picture of how things will work. In particular, they mentioned as elements to provide such a context: (1)  time- and state-behavioral details, e.g., what will be passed between the subsystems through time, the states in which operations can be performed, and how the failures/errors should be handled; (2) traceability to the original requirements; and (3) details about the technology that will be used on the production environment, so more proper tooling can be selected for the development environment early on. Regarding (1), it is worth mentioning that the lack of details about the dynamic behavior of the system was also identified as an issue yet to be addressed in the LOFAR+ case \problemref{P9}.

The quotes coded as \coderef{Life Cycle Related}{7}), on the other hand, highlighted that the initial partition of the system was mostly guided by consortium divisions (which was necessary due to the scale of the project, as described in Section~\ref{sec:SKA-overview}) rather than logical products, with the ICDs for demarcating responsibilities between the corresponding subsystems \practiceref{B3}. According to some of these respondents, this made the boundaries defined by the consortia, and the boundaries of the existing observatories incongruent (i.e., the partition was defined by responsibilities, not by the systems per se). Because of this, and in addition to the overly complicated architecture mentioned in Section~\ref{sec:design-related-practices}, this led to the emergence of unofficial, internal ICDs, that are in some cases not documented. Furthermore, it was pointed out that: (1) the ICDs were often developed by people without enough experience, and not supported by prototyping/up-front design; (2) not all parties were equally involved, hence the input level was not the same; and (3) ICDs have not been properly updated as the design progressed after the consortia phase.

\begin{figure}[t]
	\centering
	\includegraphics[width=.9\linewidth]{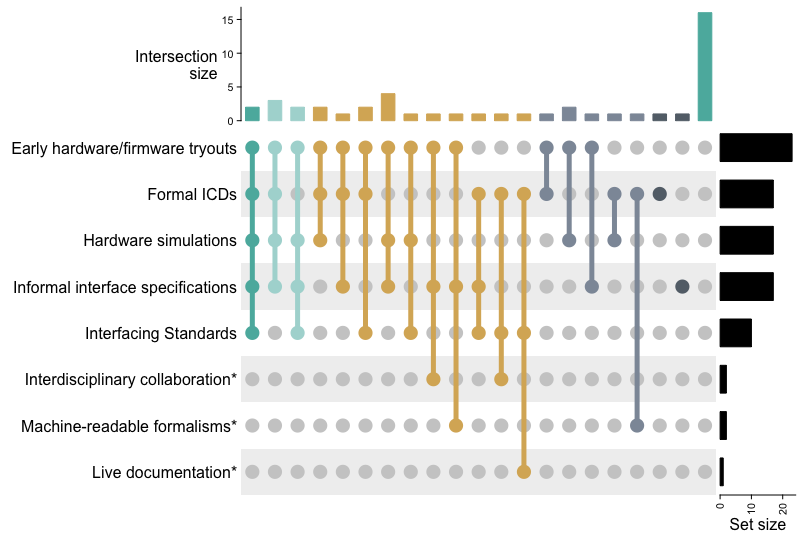}
	\caption{Responses to Q4.5 about the early integration practices adopted in the SKA to avoid or mitigate interfacing-related issues: incomplete interfaces or interfaces that do not capture mutual expectations. The new practices reported by respondents include an `*'.}
	\label{fig:45practicesmutualexp}
\end{figure}

The multiple-choice Question 4.5 concerned the \textit{practices adopted by the SKA to address the issues of interfaces that do not capture the mutual expectations of the involved parties} \problemref{P7}. Among the responses (as presented in Figure~\ref{fig:45practicesmutualexp}), which included the early-integration practices identified in the LOFAR+ case, early hardware/software tryouts \practiceref{B4} is the more recognized one. It is followed by the formal documents (ICDs), hardware simulations, and other informal interface specifications with a similar number of selections. As additional practices to address this issue, from the seven additional entries given to the `Other' option, it is worth highlighting one mentioned twice: the adoption of a semiconductor industry practice, namely \textit{the use of a machine-readable formalism as a `single source of truth' for the interfaces within the teams}, and a related workflow for the \textit{automatic generation of key components for both sides of the interface} (e.g., interfaces and documentation).  In the same spirit, having a live, constantly updated documentation (one of the goals of the aforementioned workflow), was also included as a key practice. Finally, the practice of \textit{interdisciplinary collaboration}, also added by the respondents as `Other' practice, is consistent with Figure~\ref{fig:44internalintmexp}, as hardware/software interfaces within cross-functional teams seem to have significantly fewer issues regarding capturing mutual expectations in the interfaces.

\subsection{Issues still under exploration in LOFAR+ in the context of the SKA}

\begin{figure}[t]
	\centering
	\includegraphics[width=0.75\linewidth]{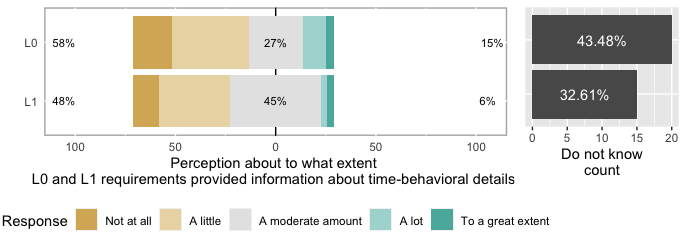}
	\caption{Distribution of the responses of Q2.5 about the perception of the extent that L0 and L1 requirements provided time-behavioral related information.}
	\label{fig:25dynbeh-req}
\end{figure}

\begin{figure}[t]
	\centering
	\includegraphics[width=0.75\linewidth]{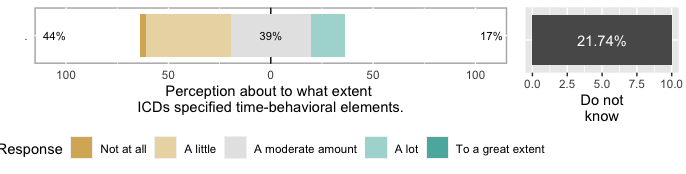}
	\caption{Distribution of the responses of Q4.3 about the extent to which ICDs captured details about the dynamic behavior of the involved subsystems.}
	\label{fig:43dynbehicd}
\end{figure}

In the LOFAR case, the \textit{lack of alignment between hardware and software development cycles} \problemref{P8}, and the \textit{missing details about the dynamic behavior of the system} \problemref{P9} in both requirements and ICDs, were identified as yet to be addressed issues that hindered the interplay between SE and SWE architecting processes (grey area in Figure~\ref{fig:basecontext}). According to the insights about problem \problemref{P9} discussed in Section \ref{sec:interfacing-pract-prob}, and the perception of the respondents on the extent to which dynamic/time-behavioral details are given in ICDs and system requirements (Figures~\ref{fig:25dynbeh-req} and~\ref{fig:43dynbehicd}), this seems to be also an important, yet un-addressed issue in the SKA, especially with respect to requirements.


\begin{figure}[t]
	\centering
	\includegraphics[width=.75\linewidth]{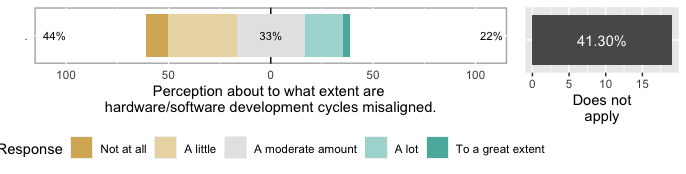}
	\caption{Distribution of the responses of Q4.6 about the extent to which ---if the respondent has been involved in hybrid hardware/software subsystems--- the development cycles were misaligned.}
	\label{fig:46misalignment}
\end{figure}

When it comes to Issue \problemref{P8}, on the other hand, the perception about the extent to which the development cycles are misaligned in the SKA is fairly positive: 77\% rated it from non-existent to only moderate, as seen in Figure~\ref{fig:46misalignment}. From the 20 responses given to the open-ended Question 4.7, most of them were categorized as \coderef{SAFe Related}{10} (see Table A.8), and provided further insights about the practices related to this perception. The responses coded with said category pointed out that hardware and software development has been aligned through SAFe's Program Increments (PIs). More specifically, it was mentioned that this has been working successfully in one of the teams\footnote{CSP Interface Prototyping Agile (CIPA) Team.} by pulling the firmware/FPGA teams into SAFe, making these components formally part of the PI's development process (in other words, mirroring these components development with the SAFe agile). Interestingly, the practice of using automated artifacts generation workflows, supported by machine-readable interface specifications (\coderef{Machine Readable Formalisms}{4} discussed in Section~\ref{sec:interfacing-pract-prob}), was also mentioned in this question by members of the same ART's team (three out of the five that participated in the study). This arguably means that this is a local practice with a seemingly positive impact on aligning the two life-cycles, and as a mean to properly capture the mutual interfacing expectations of the involved hardware/firmware/software parties. 

\section{Answers to Research Questions}\label{sec:discussion}
In the following section, we discuss the results of the study by answering each of the research questions identified in Section~\ref{sec:study-design}.

\begin{mdframed}[innertopmargin=-5pt,nobreak]
\subsection*{RQ1.1: To what extent the harmonization practices identified in LOFAR+, as applied in the context of the SKA, address the same kind of issues on requirements?}    
\end{mdframed}

For systems of the scale of SKA, \textbf{a formal SE process where the requirements were defined up-front~\practiceref{B1}, has been key not only in establishing properly defined interfaces~\problemref{P6} and requirements~\problemref{P1}, but also in preventing essential features from being missed down-the-line of the development process} (e.g., in lower level requirements and within the agile process). According to the survey results, this seems to be the case especially when working in combination with an agile process, where the understanding of the system is expected to emerge over time. However, \textbf{the results also suggest a friction between these two engineering disciplines (\textit{traditional systems engineering} and \textit{scaled agile}), which seems to be related to the said front-loaded requirements}: on the one hand, it is perceived that working with an agile approach could lead to lose the focus on the original requirements (e.g., through working by assumptions rather than requirements); on the other hand, it is perceived that the provided system-level requirements, by being long-term ones, are difficult to follow within the short-term phases of an agile process. This could explain, to some extent, why some of the consequences of having incomplete requirements identified in LOFAR+ are seemingly present also in the SKA, namely the need for making assumptions while working with them, and the little influence such requirements have on early test design. 

\begin{mdframed}[innertopmargin=-5pt]
\subsection*{RQ1.2: To what extent the harmonization practices identified in LOFAR+, as applied in the context of the SKA, address the same kind of issues on subsystem design decisions?}\label{sec:RQ12}
\end{mdframed}

There is a general perception that, given the current setting of the SKA project, in most cases local design decisions are not having a negative impact on the overall system design and qualities. Both the early partition of the system \practiceref{B3} and the hierarchical co-architecting \practiceref{B2} seemingly contributed to avoid such a negative impact~\problemref{P4}. It is therefore fair to conclude that \textbf{both practices (\practiceref{B2} and \practiceref{B3}) can be seen as confirmed, albeit to a different extent.} Although practice \practiceref{B3} has mostly a moderate to positive perception in this regard, it is seen by others as something that hindered the Solution Intent, as it led to an over-complicated architecture. However, this is arguably due to such an early partition being guided by a consortium division rather than by logical products.
In contrast, practice \practiceref{B2}, not only has a more positive perception, but also emerged (on the open-ended responses) as a key practice to address the impact of changes at the subsystem-level~\problemref{P5}. %
However, the survey results also revealed particular instances where, despite practices \practiceref{B3} and specially \practiceref{B2} being applied, local design decisions had (to a minor extent) a negative impact on the system qualities. As such instances are related to local decisions guided by the misinterpretation of the requirements, or a personal experience bias, this could be another call for the introduction of intermediate, short-term requirements as discussed for \textit{RQ1.1}. 

At the same time, further analysis of the survey results also revealed a new category of local design decisions that had a negative impact on the system architecture and qualities: \textit{workarounds to deal with higher-level, prescriptive software design decisions}. That is to say, early prescribed design decisions appear to have constrained the space for making decisions at the subsystem level, which led to workarounds to fulfill the subsystem goals. This, in turn, led to (sub-)system inefficiencies. Interestingly, the discussion of these two kinds of `negative' local design decisions (the ones related to requirements, misinterpretations or personal bias, and the ones forced by higher-level decisions), shows an apparent contradiction between: a) the participants that call for a more holistic view of the system early on for making local design decisions; and b) the participants that call for less prescriptive high-level decisions. This is seemingly another sign of friction between the two engineering perspectives combined in the SKA. Further investigation on the causes of this friction is required.

\begin{mdframed}[innertopmargin=-5pt]
\subsection*{RQ1.3: To what extent the harmonization practices identified in LOFAR+, as applied in the context of the SKA, address the same kind of issues on subsystem interfacing?}
\end{mdframed}
	
\textbf{Practices for early integration} such as \textit{harware/firmware tryouts} and \textit{hardware simulations} \practiceref{B4} \textbf{appear to be confirmed as an approach to create interfaces that fullfill the mutual expectations of their involved parties}~\problemref{P7}. Furthermore, it is worth noting that these tend to be used \textbf{in conjunction with formal ICDs and other alternative specifications} (discussed in Section~\ref{sec:rq23}) for this purpose. This is arguably related to the fact that, according to the results, there are particular subsystems where high-level interfaces, or ICDs, are too vague and incomplete (although this is seen as a deliberate decision by some). Among the elements referred as commonly missed on the ICDs were the traceability of the requirements, and the time- and state- behavioral details, which was also a recurrent theme in the LOFAR+ case. Furthermore, it is worth highlighting that the early partition of the system \practiceref{B3} emerged when exploring this research question with a negative connotation: it was perceived that this early partition, by leading to an overcomplicated architecture discussed in Section~\ref{sec:RQ12}, contributed to the emergence of obscure or not properly documented ICDs.

\begin{mdframed}[innertopmargin=-5pt]
\subsection*{RQ2 - What additional cross-disciplinary harmonization practices have been adopted by the SKA?}\label{sec:rq23}
\end{mdframed}

In the following, we present a summary of additional practices for harmonizing architecting processes in Systems Engineering and Software Engineering, as identified by the practitioners, and their relation with the cross-disciplinary issues already identified.

\subsubsection*{\practiceref{B5} Scaled agile in the context of a traditional systems engineering process.}

As discussed in RQ1.1, requirements defined up-front through a rigorous SE process are seen as key in large-scale systems like the SKA. Therefore it is not surprising, as described in the LOFAR+ case study, that unclear or incomplete requirements ---which happens to a lesser extent also in the SKA--- are among the causes of major integration or operational issues. The results of the data analysis show that \textbf{the adoption of a scaled agile framework like SAFe is seen as the key element to address unclear/incomplete requirements~\problemref{P1}}, as agility has shown to be successful when it comes to filling the gaps as the system evolves. Furthermore, \textbf{the \textit{Program Increments} of the adopted agile framework, when including the firmware/FPGAs teams, is seen as a successful approach for aligning hardware and software development cycles}, an issue that was un-addressed in the context of LOFAR+ \problemref{P8}.

\subsubsection*{\practiceref{B6} Cross-level communication and transparency}

In regard to practices that prevent the negative impact of local design decisions in the overall system~\problemref{P4}, the results suggest that \textbf{above any rigorous process, transparency and communication across different levels within the project, are key to prevent and mitigate this kind of issues}. This includes not only giving access to the right people across the organization, but also adopting strategies like the \textit{Communities of Practice} (CoP), which provides a loose structure that brings together people with common concerns, and enables cooperation across multiple domains~\cite{knaster2018safe}. Along the same lines, communication between the architects involved at both ends of the ICDs seems to be essential in order to achieve consensus-driven solutions. Overall, it is worth noting that the perception that~\problemref{P4} is an issue that happens to a minor extent on the SKA, suggests the positive influence this communication-oriented practice has had for its mitigation.

\subsubsection*{\practiceref{B7} Intermediate interfacing specifications and automation workflows}

According to the results, \textbf{the issues of lack of alignment between hardware and software development cycles~\problemref{P8} and of mutual expectations not captured by interfaces~\problemref{P7}, are being successfully addressed ---by some groups within SKA--- by means of a Model Driven-like approach.} In this approach, the low-level details of the interfaces between hardware and software (which are originally defined in human-only-readable documents, e.g., ICDs) are captured and managed as the source of truth for both parties, through a machine-readable formalism. This formalism, in turn, has allowed the development of tools and workflows for the automatic generation of key artifacts for both hardware/software parties as the resulting subsystem evolves. However, as this practice has been evidenced only in a few groups within the SKA, it is worth to be further explored and compared with its counterpart in the microcontrollers industry, from which it was inspired. In contrast, regarding issues yet to be addressed, \textbf{it is worth noting that the lack of time-behavioral details on both requirements and interfacing specifications \problemref{P9} seem to be a recurrent theme in (ultra-)large-scale systems like LOFAR+ and the SKA}.

\section[Implications]{Implications for practitioners and researchers}\label{sec:implications}

The following  summarizes our understanding of the implications of this study's results for practitioners and researchers. 

\subsection{Agile and Systems Engineering principles in the context of software-intensive SoS}

The plan-driven methods prescribed by the traditional SE principles have been historically shown to provide a logical process structure for large-scale, multidisciplinary SoS projects~\cite{haberfellner2019systems}. However, when the projects are software-intensive, such plan-driven approaches have been often criticized, as they make the development of said software elements difficult~\cite{carson20134}. This is particularly due to the assumption that changes in software, like in hardware and mechanical systems, will decrease over time~\cite{santander2017agile}. For this reason, the relationship between the Agile and Systems Engineering principles has been extensively explored, leading to multiple approaches for mixing them, such as the ones supported by \textit{Model-based Systems Engineering}, where agile principles are applied on top of evolving, high-precision system models~\cite{muscarella2020systems, douglass2015agile}. 

The SKA, on the other hand, and arguably due to its scale, has followed a tailored approach that combines both sets of principles by following a traditional plan-driven design, and then, based on the defined requirements and interfaces, transitioning to a scaled agile process. The results of our study highlight the importance of system requirements properly defined up-front in this particular and similar approaches, as they capture key features that cannot be missed down the line in the agile process, and that will not necessarily arise as part of the emergent understanding of the system. However, these requirements seem to be an important factor of friction between the SE perspective of a system, and the one from the agile teams working on the software-heavy elements of it, due to its long-term nature. For this reason, practitioners in this context should take into consideration the importance of (1) \textbf{deriving intermediate (short-term) requirements from the long-term ones, so that agile teams are less prone to lose the focus on them}, and (2) \textbf{ensure and continuously validate the traceability between long-term and short-term requirements}.

\subsection{System decomposition in large-scale, directed SoS}

Decomposing complex, large-scale systems (of systems) into subsystems through e.g.~logical decomposition is a core SE practice, which, with the use of formal interfacing specifications (e.g., ICDs) defines clear boundaries that allow multiple teams to work concurrently on the system design and development. In the LOFAR+ case, the transition from a layer-based decomposition (hardware/firmware/software) to a subsystem-based one, showed that this SE practice helped in preventing local design decisions from having a negative impact on the overall architecture and system qualities. Likewise, on the SKA project, this practice was key in delegating the design of such a massive system to multiple consortia. However, as this decomposition was guided by delegation of responsibilities rather than logical products, which had adverse implications on the system qualities; this suggests the need for further \textbf{research on reconciling functional decomposition with organizational decomposition at this scale}. Furthermore, the communication-oriented practices adopted by the SKA regarding ICDs, that is to say, the work towards consensus-driven solutions with the architects involved at both ends of the ICDs, calls for further \textbf{research on how to achieve this communication effectively, given the potentially very different knowledge domains involved}.

\subsection{Driving and guarding the system qualities}

In the LOFAR+ case study multiple scenarios were identified where subsystem-level design decisions (e.g.~when boundaries between subsystems were not clearly demarcated) had a negative impact on the overall system architecture and its qualities. In the SKA, seemingly due to the co-architecting roles and the overall scaled agile process already in place, this study revealed other kind of design decisions that, indirectly, also have a negative impact; admittedly, this happens to a minor extent. High-level, prescriptive, software-related technical decisions (e.g., algorithms parameters, protocols selection, abstraction layers, etc.) lead practitioners to resort to workarounds on the subsystems implementation to fulfill their goals. These workarounds however end up hindering the overall system qualities. This calls practitioner to give \textbf{higher priority to the negotiation of decisions with a cross-cutting impact on multiple agile teams in the context of a scaled agile framework}. Furthermore, the apparent contradiction between the engineers that call for a more holistic view of the system to make better informed decisions, and those that call for less high-level prescriptive specifications for the system calls for further \textbf{research on practices for reconciling the bottom-up and top-down design approaches that coexist in systems of this scale}.

\subsection{Subsystems interfacing}

This study corroborated the importance of formal interface specifications as enablers of effective collaboration between cross-disciplinary teams~\cite{Muscarella2020}. However, it also confirmed that the lack of details about the dynamic aspects of the interfaces (e.g., time and state-behavioral details) is a recurring phenomenon that is seemingly causing problems (mostly) on the software side of the involved subsystems. This suggests the need for further \textbf{research on formalisms or Domain Specific Languages (DSLs) for the specification of such elements}, especially considering the cross-disciplinary settings of large-scale systems like LOFAR+ and the SKA and the differences in the terminology they exhibit, as discussed by previous studies in SE-SWE interfacing~\cite{sheard2019systems}.

At the same time, the study further revealed what seems to be a promising best practice ---given the perceived success within the team that adopted it--- for aligning the life cycles of hardware and software teams, as well as managing the lower-level interfaces of such teams: the use of machine-readable interfacing specifications as a single source of truth for both hardware and software teams, and workflows for the generation of relevant artifacts for both parties. Although this practice is well known in the semi-conductors domain, its success in the context of large-scale systems calls for further \textbf{research on how to make it consistent with the higher-level interfaces (i.e., ICDs) that are often defined in human-only readable formats}, and from which the lower-level interfaces are actually meant to be derived.

\section{Threats to validity}\label{sec:ttv}
In the following we discuss potential threats to the validity of this study, and the steps taken to mitigate them. 
We use Runeson and H\"ost~\cite{runeson2009guidelines} as a guide for this purpose. \textit{Internal validity} does not apply as this study does not examine causalities (\cite{yin2017case}).

\paragraph*{Construct validity}

Construct validity refers to the degree the research instruments, in this case the online survey, are consistent with the research questions, and reflect what the researchers have in mind while designing them. To improve the construct validity of the study, as discussed in Section~\ref{sec:study-design}, the survey was piloted by three engineers with experience in the case subjects of this and the previous study (the SKA and LOFAR+ projects). They provided feedback about the wording and the terminology used in the questions. The survey was also reviewed prior to its promotion to the target population, by one of the co-authors, who also has experience in the application domain, as well as one of the SKA architects. 

To prevent biasing the perception of the participants about the results of the previous study, we designed the survey without any presumption of expectation of familiarity with the LOFAR+ case. To do so, as described in Section~\ref{sec:study-design}, we first explored how such practices have been applied in the SKA, so that the survey questions can be designed using SKA-specific context and terminology. Furthermore, and to avoid bias by leading questions and make the confirmation more reliable, the survey did not use close-ended questions to directly gauge the relationship between specific practices and issues (e.g., \textit{to what extent practice A improves issue X?}). Instead, such relations were mostly evaluated through the themes (or codes) that emerged from the open-ended responses to questions of the type \textit{how has the SKA prevented or mitigated X?}, and complemented by the insights given by the closed-ended ones. Independent verification of this process is possible through the replication package of this study. 

Another construct-related issue that can affect negatively the confirmatory robustness of our study is due to the risk of the number of opted-out questions per role (described in Section~\ref{sec:demographics}) skewing the data. Likewise, this threat was mitigated, to some extent, by accompanying them with related open-ended questions.
Ultimately, we believe that the risk due to opt-outs is justified when compared to the ones created by forcing participants to provide an answer to a question that they are not comfortable or familiar with. Doing so would either introduce noise in the replies due to participants choosing the middle/neutral option as the default response, or create mid-survey drop-out issues with frustrated participants.

\paragraph*{External validity}
External validity refers to the extent research findings can be generalized from the sample to the target population. In this case, external validity can be seen from two viewpoints: (1) the SKA project as a sample of all the directed SoS, and (2) the study participants as a sample of total population of the engineers involved on the SKA's ART. With respect to the former, although this study is conducted in the same application domain (radio-astronomy scientific instruments) as the one it aims to confirm and extend, we believe that its characteristics make it still representative of large-scale, directed SoS. However, validation of the findings in other engineering domains remains part of our future work. Regarding the latter, this study was subject to self-selection bias due to its non-probabilistic sampling ---the participation to the study was based on an open call to the target population. However, as described in Section~\ref{sec:demographics}, we find the sample representative enough not only of the roles (considering that all of them are involved on the ART), but also on their affiliations and distribution across the globe. The number of participants was relatively limited (46), but this was due to the very specific target population of the engineers involved in the scaled-agile part project.

\paragraph*{Reliability}

Reliability refers to the extent that the data and its analysis are dependent on the researchers that conducted the study. To mitigate researcher bias in the process, two of the authors were involved in the quantitative data analysis, and came to an agreement with a third author about its interpretation given the results of the qualitative data analysis. The qualitative data analysis, in turn, was conducted by the first author, and its outcome validated by other two authors. Furthermore, both qualitative and quantitative analysis can be verified through the replication package provided for this purpose.

\section{Conclusions}\label{sec:conclusions}
The concept of Systems of Systems has gained significant attention from multiple engineering disciplines, as a powerful concept to deal with the complexity of modern systems and the goals they aim to attain. This is particularly the case for Systems Engineering (SE) and Software Engineering (SWE), two disciplines that are deeply intertwined in the development of many engineered\footnote{In contrast to naturally-occurring ones.} SoS: the former managing the system complexity and the disciplines involved in it (including SWE), and the latter providing key features through the software-heavy components scattered across it. Despite this mutual dependency, however, there is little research on how to harmonize these disciplines when working together, in particular regarding their respective architecting processes. This study ---the third one in a series of related studies--- contributed to closing this gap by exploring the architecting harmonization practices previously identified in the LOFAR+ case when applied in the context of the SKA. Based on its results, it evidently confirms the findings of the LOFAR+ study with respect to such practices. Furthermore, it unveiled additional best practices and provided further insights to be considered by practitioners when architecting SoS, and to be explored by researchers in the area.

In particular, this study  shows that an agile process is indeed a good complement for traditional SE processes when it comes to  incomplete/vague system requirements, something arguably unavoidable in systems with the scale of the SKA. Likewise, strategies that enable discussion on common concerns across the systems, and custom interfacing management approaches, are further examples of practices that complement the SE practices (e.g., the \textit{critical design reviews}) in preventing the negative impact of local design decisions, and aligning development cycles, respectively. The areas of improvement identified in this case study, on the other hand, provide important insights to be considered in future projects and in the aforementioned research agenda.  For instance, this study suggests that having short-term intermediate requirements (derived from the long-term ones) would improve the alignment between the system originally envisioned on the early stages of the project (when such requirements are defined), and the system understanding that emerges from the agile process. When it comes to the early partition of the system, and their demarcation with ICDs, this study highlights the importance of further attention ---and further research--- on (1) system breakdown techniques that better align functional decomposition with large-scale responsibilities distribution, and (2) the way software-specific design decisions, with impact across multiple agile teams are made, as a measure to prevent architectural smells. 

Overall, using the SKA has exemplified how, with the right practices rather than a `clash of cultures', disciplines can complement each other while combining the best of both worlds. This makes us believe that this work can be seen as the baseline for a new direction in the current SWE research agenda in SoS architecting: \textit{well-informed practicing of SWE in the context of SoS projects governed by a SE process}. 

Along this line, as a future work we plan to build on the findings and work towards a contribution to some of the prevalent issues identified by this and previous studies, and particularly the communication-related ones, concerning terminology, completeness, and consistency. For this purpose, we plan to devise processes and artifacts for the management of cross-disciplinary interfaces that work under some of the aforementioned practice-related principles, namely transparency and consensus-oriented communication, in combination with the automation possibilities of modern technical documentation, whose potential has been shown in this and other domains in the literature.

\section*{Acknowledgments}
	The authors would like to thank the members of the SKA organization and the engineers involved in the project for their time in helping us conduct this research. Special thanks to Nick Rees, Juande Santander-Vela and Marco Bartolini for their support and feedback while conducting and promoting this study within the 10th PI-planning event. We would also like to particularly thank the participants of the survey pilot for their invaluable input: Marcel Loose, Stefan Wijnholds, and \'{A}gnes Mika.
	
\clearpage

\bibliographystyle{unsrt}
\bibliography{refs,dcs,sos-patt-archs,sec-studies,iface-mgmt,icdm-core-refs,mbse-interfaces,gold-refs,general,sos-examples,base-refs,ska-refs}

\newpage

\appendix
\section[Sample Quotes]{Sample quotes of the open questions coding process}\label{sec:appendix}
\subsection{Requirements-related questions}

\footnotesize
\setcounter{table}{0}

\begin{longtable}{@{}m{0.22\textwidth} m{0.65\textwidth} c@{}}
	
	\caption{Prominent codes that emerged from the \textit{Qualitative Content Analysis} process on the responses to Question 2.6 (favorable views)}\\
	
	\toprule  
	\multicolumn{1}{c}{\bfseries Code}&
	\multicolumn{1}{c}{\bfseries Sample Quotes}&
	\multicolumn{1}{c}{\bfseries Occurrences}\\ \midrule
	\endfirsthead

	\textsc{SOLID-GROUND}   &  \scriptsize{\textit{There has to be a point of reference, but these are never complete. With intelligence and good will, the gaps were overcome or filled in as necessary.},  \textit{Neutral, as it was a necessary evil because of the early partitioning of the project.}} &   7       \\
	\hline
	\textsc{SOLUTION-INTENT-BASE} & \scriptsize{\textit{In almost all cases, the the L1/L2 requirements have helped not hinder the solution intent provided use case summaries additionally exist. Defined up-front details regarding edge cases, data formats and the nice-to-have requirements additionally help define the solution intent.}, \textit{L1/L2 requirements help define solution intent.}}  & 4 \\
	\hline
	\textsc{HELPFUL-IN-GENERAL} & \scriptsize{\textit{They help immensely}, \textit{They certainly provide guidance on the overall framework of requirements. It is clear that L1 and L2 requirements will evolve with time. The SAFe approach should help mitigate risks with changing requirements.}, \textit{I think this helps, especially wrt hardware and FPGA development, which is code but not "software".}}
		 & 10 \\
	\hline
	\label{tab:quotes-q2-6p}
\end{longtable}
\begin{longtable}{@{}m{0.22\textwidth} m{0.65\textwidth} c@{}}
	
	\caption{Prominent codes that emerged from the \textit{Qualitative Content Analysis} process on the responses to Question 2.6 (negative views)}\\
	
	\toprule  
	\multicolumn{1}{c}{\bfseries Code}&
	\multicolumn{1}{c}{\bfseries Sample Quotes}&
	\multicolumn{1}{c}{\bfseries Occurrences}\\ \midrule
	\endfirsthead

	\textsc{LITTLE-WEIGHT}   &  \scriptsize{\textit{`Not much. I don't see these requirements having a big weight in the decisions for OMC. I do see them being a main point in the SRCs though.'},  \textit{`The inform it but not more than that.'}} &   2   
	
	\\
	\hline
	\textsc{NOT-STRICTLY-FOLLOWED} & \scriptsize{\textit{`I think the current SKA solution intent was driven mainly by assumptions, experience of the architects rather than the L1/L2 requirements are they were written. The L1/L2 requirements are currently being almost "shelved" in my opinion and are not referenced often enough.'}, \textit{`It would help, if we stick to those requirements. But budget / time considerations typically means we have to work with what we have when the times comes to implement things.'}} & 3 \\ \hline		
	\label{tab:quotes-q2-6-n}
\end{longtable}

\begin{longtable}{@{}m{0.22\textwidth} m{0.65\textwidth} c@{}}
	
	\caption{Prominent codes that emerged from the \textit{Qualitative Content Analysis} process on the responses to Question 2.6 (additional issues)}\\
	
	\toprule  
	\multicolumn{1}{c}{\bfseries Code}&
	\multicolumn{1}{c}{\bfseries Sample Quotes}&
	\multicolumn{1}{c}{\bfseries Occurrences}\\ \midrule
	\endfirsthead
	\caption{Prominent codes that emerged from the \textit{Qualitative Content Analysis} process on the responses to question 2.6 (additional issues) --- continued from previous page}\\
	\toprule  
	\multicolumn{1}{c}{\bfseries Code}&
	\multicolumn{1}{c}{\bfseries Sample Quotes}&
	\multicolumn{1}{c}{\bfseries Occurrences}\\
	
	\endhead
	
	\textsc{TRANSITION-TO-AGILE-ISSUES}   & \scriptsize{\textit{`Depends on which 'level' your scope is focussed on but a clearer link between tasks and the Lx requirements they are helping to support would be good'},	\textit{`I think our requirements are well defined but I also think the link between them and the feature on which teams are working on isn't clear. This makes it a bit difficult to understand if the emergent understanding is deviating from what we originally thought'}}.
	&   5      \\
	\hline
	\textsc{MISSING-INTERMEDIATE-ELEMENTS} & \scriptsize{\textit{`The L1/L2 are clearly useful but the main issue I see is that the requirements were/are based on the final solution which will take many years to build. What is missing imo are specific requirements per Array release. These release requirements should then drive the architecture for that release.'}, \textit{`The L1/L2 requirements specify the final SKA telescope systems, but not intermediate versions such the early array assemblies (AA0.5, AA1, etc.). They are also focussed on the steady-state use of the telescopes for observing projects, so they do not necessarily capture the activities that need to take place during construction and commissioning. The lack of short- and medium-term requirements has hindered the development of the solution intent, especially the specification of tests for the prototype software systems.'}} & 6 \\
	\hline
	\textsc{IMPERFECT-REQS-ISSUES} & \scriptsize{\textit{`Lack of clear and well defined L2 and L3 requirements for software dominated sub-systems  is hindering progress and results in constant re-factoring.  Sub-systems where L2 and L3 requirements are well-defined do not experience so much churn.'}, \textit{`I think they help, but they were imperfect in the first place and they would have helped more had they been more complete.  There was a lack of iteration to ensure that the flowed-down requirements were complete at the L3 level (and hence all the way back up through L2 and L1).  In some cases, there were obvious gaps discovered at L3, which highlight missing L2 and/or L1 requirements, and this up and down process should have been repeated several times to shake out the whole requirements tree.'}, \textit{Help, though not defined well.}} & 3 \\
	\hline
	
	\label{tab:quotes-q2-6-int}
\end{longtable}

\begin{longtable}{@{}m{0.22\textwidth} m{0.65\textwidth} c@{}}
	
	\caption{Prominent codes that emerged from the \textit{Qualitative Content Analysis} process on the responses to Question 2.7}\\
	
	\toprule  
	\multicolumn{1}{c}{\bfseries Code}&
	\multicolumn{1}{c}{\bfseries Sample Quotes}&
	\multicolumn{1}{c}{\bfseries Occurrences}\\ \midrule
	\endfirsthead
	\caption{Prominent codes that emerged from the \textit{Qualitative Content Analysis} process on the responses to Question 2.7 --- continued from previous page}\\
	\toprule  
	\multicolumn{1}{c}{\bfseries Code}&
	\multicolumn{1}{c}{\bfseries Sample Quotes}&
	\multicolumn{1}{c}{\bfseries Occurrences}\\
	
	\endhead
	
	\textsc{PROTOTYPES-TESTING}  & \scriptsize{\textit{'Involving stakeholders, by integrating and demonstrating functionality as quickly as possible. Unclear requirements aren't really as much of a problem when you can just get feedback from the person that wrote them directly'}, \textit{'Most requirements are incomplete not because the issues involved are being missed, but because additional clarity can only come with more testing of prototypes. The testing and prototyping work is helping mitigate the issues with the requirements'}, \textit{'Some of the issues were quite obvious when the Assembly Integration and Verification team were writing the acceptance tests for individual requirements. The policy of test-driven design should help to identify problems in the requirements as construction progresses.'}}
	&   5      \\
	\hline
	\textsc{AGILE-FILLING-GAPS} & \scriptsize{\textit{`Some have been mitigated against by the agile processes, however new incomplete requirements are not prevented.'}
		,
		\textit{`They certainly provide guidance on the overall framework of requirements. It is clear that L1 and L2 requirements will evolve with time. The SAFe approach should help mitigate risks with changing requirements.'}, \textit{`In Software development requirements are always changing and evolving. The initial L1/L2 requirements should be used only as a guideline. The use of agile/lean methodology helps to mitigate and cope with the inevitable changing of requirements.'}, \textit{`The SAFe solution and programme teams have recognised the lack of requirements for the early array assemblies, and are in the process of developing them in the form of use cases.'}} & 9 \\
	\hline
	\textsc{REQS-REVIEWS} & \scriptsize{\textit{'We had several reviews, teams documenting assumptions, and several requirement clarification processes'.},  \textit{'Happening to a minor extent in requirements reviews'.},	\textit{'There would have been far fewer problems had the rationale fields been populated with the reasoning that justified the requirement (and more importantly the vaues) when the requirements were developed. Some of this has been recovered during requirements reviews but often the only source is personal recollection'.}} & 4 \\
	\hline
	\label{tab:quotes-q2-7-gaps}
\end{longtable}

\subsection{Subsystem design decisions-related questions}

\begin{longtable}{@{}m{0.22\textwidth} m{0.65\textwidth} c@{}}
	
	\caption{Prominent codes that emerged from the \textit{Qualitative Content Analysis} process on the responses to Question 3.2}\\
	
	\toprule  
	\multicolumn{1}{c}{\bfseries Code}&
	\multicolumn{1}{c}{\bfseries Sample Quotes}&
	\multicolumn{1}{c}{\bfseries Occurrences}\\ \midrule
	\endfirsthead
	\caption{Prominent codes that emerged from the \textit{Qualitative Content Analysis} process on the responses to Question 3.2 --- continued from previous page}\\
	\toprule  
	\multicolumn{1}{c}{\bfseries Code}&
	\multicolumn{1}{c}{\bfseries Sample Quotes}&
	\multicolumn{1}{c}{\bfseries Occurrences}\\
	
	\endhead

	\textsc{HIGH-LEVEL-DESIGN-DECISIONS}  & \scriptsize{\textit{`Selection of filter coefficients for beamformers is unnecessarily prescriptive, removing the possibility of optimising the total signal chain performance with downstream components.'},
		\textit{`The required resolution of the visibilites was/is specified as single precision floating point, where only 11b integers are required given the signal to noise ratio at the receptor, and relatively sort integration time, and narrow channels.},\textit{`The transport of the visibility data using RDMA technology (from 2011) instead of UDP (from 1970s). RDMA slightly higher protocol that transports messages (visibility sets) using hardware acceleration, instead of using UDP and CPUs to reassemble visibility sets (SPEAD).'},\textit{`Choice of SPEAD protocol was premature and without performance evaluation.'}, \textit{`EF was a good idea for separating computing and scheduling.  However, there was a big computing overhead price to pay.'}, \textit{`The reliance of measurement sets and lack of an alternative strict data model or model view sometimes makes it challenging to produce robust solutions that can be tested on real data from existing antenna arrays, such as detecting autocorrelations, incomplete data, non supported flagging, and number of baselines in the data.'}, \textit{`Choice of deployment infrastructure (k8s, helm) that might be suitable for TMC side but not necessarily appropriate for SDP (HPC environments) without a proper technical evaluation.'}, \textit{`decisions on locating delay calculations. Doing it at the TMC level means that they cannot be supported at the highest rate that may be required by the CBFs especially Low. So we have to resort to using polynomial fits etc.'}, \textit{`The 'siloed' architecture from the consortia phase has led to an overly-complicated design for the control system when the subsystems were combined. The control system is implemented using Tango devices, and there are redundant devices that do not provide additional functionality.'}, \textit{`The use of base-classes has meant a number of sub-systems have been reliant. Changes to these base-classes have necessitated updating of a number of subsystems and led to broken deployments whilst changing.'}}
	&   8      \\
	\hline
	\textsc{LOCAL-INTER-PRETATIONS} & \scriptsize{`\textit{Some decisions that made the correlators easier to design, from previous knowledge of the team, had an impact on the availability and engineering maintenance of them. The review of those decisions has now provided better correlators for both telescopes, which are also now more resilient}'.
		`\textit{Assumptions on how subarrays are to be used}'.
		`\textit{I think the issue here is that it's only now that we are looking at the telescope in an holistic way. We see each element has designed their systems in the way that makes the most sense to them. However when you connect these systems together and look at the system as a whole it may no longer make as much sense. In the area of networks , many subsystems have network elements inbuilt into them but only now are we beginning to look at the network as a whole. If this had been investigated earlier in the design lifecycle with some design decisions made up front I believe we would have saved a lot of time and effort.}'
		\textit{`In a couple of cases local interpretation of a capability requirement (number of beams, independence of behaviour between sub-arrays) has lead to reduced flexibility at the system level which would have made it difficult to use the system efficiently. So far, we have been able to revise these in a constructive way, but it has involved personal intervention to encourage re-thinking.'}, \textit{`Impact of changes in design in a subsystem have rippling effect on documentation and other written artefacts which take time to 'catch up' and can mean some disconnect until changes are reflected 'everywhere''}, \textit{`I am not aware of an example where decisions made by one sub-system negatively affected other sub-systems. But I am aware of the cases where lack of decisions (undefined requirements and architecture) in some areas affected other sub-systems and overall progress.'},\textit{`the problem is mostly that of a lack of design decisions, either because of a fear of obsolesence when deciding too early or in order to characterize all possible solutions'}, \textit{`the lack of design choices being made leaves little scope for mitigation efforts'} } & 8 \\ \hline
	
		\textsc{INTERFACING/ INTEGRATION-RELATED} & \scriptsize{\textit{`This is early days, so we are talking fairly banal qualities here - primarily just overall stability, some performance.'}, \textit{`Something like the current troubles with getting Tango to work predictably affect basically everybody on some level.'}, \textit{`You could say that we are struggeling with implementing interfaces effectively on both sides.'}, \textit{`Minor issues around interface harmonisation'}, \textit{`OET-TMC-LMC integration has been difficult and slow affecting testability, and stabilty.'}, \textit{`Interface definitions don't make sense to parties on either side of the interface'}} & 4 \\
	
	\hline
	
	\label{tab:quotes-q3-2-dd}
\end{longtable}

\begin{longtable}{@{}m{0.22\textwidth} m{0.65\textwidth} c@{}}
	
	\caption{Prominent codes that emerged from the \textit{Qualitative Content Analysis} process on the responses to Question 3.6}\\
	
	\toprule  
	\multicolumn{1}{c}{\bfseries Code}&
	\multicolumn{1}{c}{\bfseries Sample Quotes}&
	\multicolumn{1}{c}{\bfseries Occurrences}\\ \midrule
	\endfirsthead
	\caption{Prominent codes that emerged from the \textit{Qualitative Content Analysis} process on the responses to Question 3.6 --- continued from previous page}\\
	\toprule  
	\multicolumn{1}{c}{\bfseries Code}&
	\multicolumn{1}{c}{\bfseries Sample Quotes}&
	\multicolumn{1}{c}{\bfseries Occurrences}\\
	
	\endhead

	\textsc{COMMUNICATION -RELATED} & \scriptsize{\textit{`We have a much higher level of transparency, and access to the relevant people, which makes it easier to prevent this issues.},
		\textit{`Early integrations; better communication of solution level'}
		\textit{`Communication within the architecture group and CoP, mostly / hopefully.'}
		\textit{`The architects are the design authority and have a lot of regular connection with the SAFe Teams, resolving ambiguities in the original design, and shepherding implementation inconsistencies as they arise.'}
		\textit{`I see it as a ongoing process. We are having discussion at several levels (formal and informal) that aim to address these issues.'}
		\textit{`Reduced by active communication and raising of concerns in planning and sprint meetings.'}
		\textit{`Better documentation of rationale of decisions and communication/review by System Engineers/Solution Architect'}
		\textit{`I think alignment and communication across levels are key. We are working for improving it.'}
		\textit{`This does vary, in some areas this has been prevented by full consideration of the potential impact when decisions are made and this works best when input is given from all levels of the project.'} }
	&   6      \\

	\hline
	
	\label{tab:quotes-q3-6-dd-prac}
\end{longtable}

\subsection{Subsystems interfaces-related questions}

\begin{longtable}{@{}m{0.22\textwidth} m{0.65\textwidth} c@{}}
	
	\caption{Prominent codes that emerged from the \textit{Qualitative Content Analysis} process on the responses to Question 4.2}\\
	
	\toprule  
	\multicolumn{1}{c}{\bfseries Code}&
	\multicolumn{1}{c}{\bfseries Sample Quotes}&
	\multicolumn{1}{c}{\bfseries Occurrences}\\ \midrule
	\endfirsthead
	\caption{Prominent codes that emerged from the \textit{Qualitative Content Analysis} process on the responses to Question 4.2 --- continued from previous page}\\
	\toprule  
	\multicolumn{1}{c}{\bfseries Code}&
	\multicolumn{1}{c}{\bfseries Sample Quotes}&
	\multicolumn{1}{c}{\bfseries Occurrences}\\
	
	\endhead
	
	\textsc{INCOMPLETE-CASES} & \scriptsize{\textit{`The TM-LFAA ICD is one example where the mutual expectations were not fully captured.'},
		\textit{`MCCS was a key one missed'},
		\textit{`ICDs seem to be fairly complete and sufficient to proceed development against.  The notable exception seems to be the monitor and control ICDs (TMC to Elements, CSP LMC to CSP sub-elements) which seem to be very vague and incomplete.'},
		\textit{`The experience has been that important details of the interface have been omitted, which only becomes apparent when you try to implement it.'}} & 4 \\ \hline
	\textsc{MISSING-ELEMENTS} & \scriptsize{\textit{`Time-behavioural elements were often not clear e.g. in what device states may a command/function be called? Handling of failures and error was not well covered.'},
		\textit{`Many details on how different components were supposed to come together'},
		\textit{`The technology used to deliver the end result. We are using tooling that clearly doesn't fit into a production environment and that will have to be addressed down the road.'},
		\textit{`Understanding of the overall requirements and system.'}
		\textit{`ICDs were structured at Element to Element level rather than Product to Product, sometimes making it difficult to track interface requirements.'},
		\textit{`Failure modes and their recovery.'},
		\textit{`No use cases/examples. No end-to-end description of how things would work.'},\textit{`Lack of context.'},
		\textit{`Details of what will be past back and forth between subsystems in real time.'},
		\textit{`Better description of software interfaces via API
		Interaction and cross-element review of the ICDs'}}
	&   10      \\ \hline
	\textsc{LIFE-CYCLE-RELATED} & \scriptsize{\textit{`Not all parties were equally involved, i.e. the effort invested in different sub-systems varied greatly, and in some cases one of the interfacing parties did not provide input.'},
		\textit{In order to pragmatically move the design forward, the design was carved up along consortia lines, and ICDs were used as contracts or demarcations of responsibility carved up between systems.  This while being necessary, has meant that the deeper downstream and upstream implications of design decisions have not always been harmonised.},
		\textit{The ICDs (especially the software sections of the ICDs) needed updating as the design progressed. This was not managed well after the end of the consortia era.},
		\textit{it seems the ICDs were kept deliberately vague; probably because it was perceived unrealsitic to advance the detailed design sufficiently and secondly, consortia boundaries were not congruent with system boundaries of existing observatories but rather seemed to reflect an ambition of how to partition responsibilities and not the systems},
		\textit{Nothing was missed as such, just ICDs were oftern 1) developed by people with not enough domain or architectural experience to be useful in several cases 2) not supported by prototyping and pen and paper up front design never really works!}
		\textit{individual ICDs for a given element were frequently negotiated by different people, leading to internal inconsistencies. The ICDs were split on consortium divisions, not logical products, leading to nedlessly complexity in some cases, and obscuring internal ICDs (which in some cases were not even documented).},\textit{Experts on existing implementations.}} & 7 \\

	\hline
	
	\label{tab:quotes-q4-2}
\end{longtable}

\subsection{Non-addressed problems related questions}

\begin{longtable}{@{}m{0.22\textwidth} m{0.65\textwidth} c@{}}
	
	\caption{Prominent codes that emerged from the \textit{Qualitative Content Analysis} process on the responses to Question 4.7}\\
	
	\toprule  
	\multicolumn{1}{c}{\bfseries Code}&
	\multicolumn{1}{c}{\bfseries Sample Quotes}&
	\multicolumn{1}{c}{\bfseries Occurrences}\\ \midrule
	\endfirsthead
	\caption{Prominent codes that emerged from the \textit{Qualitative Content Analysis} process on the responses to Question 4.7 --- continued from previous page}\\
	\toprule  
	\multicolumn{1}{c}{\bfseries Code}&
	\multicolumn{1}{c}{\bfseries Sample Quotes}&
	\multicolumn{1}{c}{\bfseries Occurrences}\\
	
	\endhead

	\textsc{SAFE-FRAMEWORK-RELATED} & \scriptsize{
		\textit{`Make firmware standards part of the fundamental software standards; make firmware and FPGA development part of SAFe.'},
\textit{`Misalignment was experienced with hardware process, the team lead (PO) managed the hardware development and provided input required for planning software and firmware development.'},
\textit{`SAFe'},
\textit{`SAFe'},
\textit{`WE have mirrored Hardware and Firmware with the SAFe Agile teams style planning
This alignment has been all within the CIPA team, so we have managed it fairly well.'},
\textit{`Close collaboration and software-firmware co-development within the same team. }\textit{`Cross-training of some team members on software and firmware.'},
\textit{`Program Increments are meant to do this, but haven't been working well. Lots of things have been late, or never done.'},
\textit{`Pulling the firmware teams into SaFE has been successful, once they learned (or taught us) the language they understood and we learned together how to break down and describe the work blocks. We still have a bit of friction over "done"-ness, but that is improving as both sides better understand the module artifacts and test strategies.'},
\textit{`PI planning and Weekly SoS are helping in this regard.'}}
	&   10      \\ \hline
	
	\textsc{MACHINE-READABLE-FORMALISMS} & \scriptsize{\textit{`...integration tests and formal interface specifications (e.g. Tango/JSON schemas).'},
		\textit{`CIPA has created a "single source" (json) description of each interface between software and the FPGA IP blocks from which code for both sides is generated by a tool. This makes changes to the interface cheaper and faster to implement.'},
		\textit{`We use JSON files backed by a well-defined schema as a single-source machine-readable document which is used to generate firmware interfaces, software interfaces, human readable documentation, and other artifacts. This is a standard practice in the semiconductor industry and we have developed a workflow around this concept using tools familiar to us.'},
		\textit{`Cross-training of some team members on software and firmware. Tools for auto-generating software from the firmware register definition (JSON file)'}} & 4 \\
	
	\hline
	
	\label{tab:quotes-q4-7}
\end{longtable}

\end{document}